\documentclass[prb,aps,onecolumn,amsmath,nofootinbib,superscriptaddress,10pt]{revtex4-2}
\usepackage[utf8]{inputenc}
\usepackage{amssymb,amsmath}
\usepackage{graphicx}
\usepackage{float}
\usepackage{natbib}
\usepackage{multirow}
\usepackage{physics}
\usepackage{subcaption}
\usepackage{natbib}
\usepackage{cancel}
\usepackage{comment}
\usepackage{xcolor}
\usepackage{booktabs}

\usepackage[final]{hyperref} 
\hypersetup{
	colorlinks=true,       
	linkcolor=red,        
	citecolor=red,        
	filecolor=magenta,     
	urlcolor=blue         
}

\usepackage{color} 

\definecolor{darkgreen}{rgb}{0,0.7,0}  
\definecolor{orange}{RGB}{255,127,0}

\newcommand\be{\begin{equation}}
\newcommand\ee{\end{equation}}
\newcommand\eps{\epsilon}
\renewcommand\th{\theta}
\newcommand\lam{\lambda}

\renewcommand{\dd}{{\rm d}}

\def\bes{\begin{subequations}}
\def\esu{\end{subequations}}

\newcommand{\cumulantplot}[2]{
\begin{figure}[H]
    \centering
    
    \begin{subfigure}{0.45\textwidth}
        \centering
        \includegraphics[height=4cm]{figures/cumulants/#1_T_0.01_alpha_pi_over_6.pdf}
        \caption{$T=0.01$, $\alpha=\pi/6$}
    \end{subfigure}
    \hfil
    \begin{subfigure}{0.45\textwidth}
        \centering
        \includegraphics[height=4cm]{figures/cumulants/#1_xi__0,3,2__alpha_pi_over_3.pdf}
        \caption{$\xi=\frac{1}{3+1/2}$, $\alpha=\pi/3$}
    \end{subfigure}

    \begin{subfigure}{0.32\textwidth}
        \centering
        \includegraphics[height=3cm]{figures/cumulants/#1_T_0.01_xi__0,3,2_.pdf}
        \caption{$T=0.01$, $\xi=\frac{1}{3+1/2}$}
    \end{subfigure}
    \hfil
    \begin{subfigure}{0.32\textwidth}
        \centering
        \includegraphics[height=3cm]{figures/cumulants/#1_T_0.5_xi__0,3_.pdf}
        \caption{$T=0.5$, $\xi=1/3$}
    \end{subfigure}
    \hfil
    \begin{subfigure}{0.32\textwidth}
        \centering
        \includegraphics[height=3cm]{figures/cumulants/#1_T_2.0_xi__0,3_.pdf}
        \caption{$T=2.0$, $\xi=1/3$}
    \end{subfigure}

    \begin{subfigure}{0.24\textwidth}
        \centering
        \includegraphics[width=1\linewidth]{figures/cumulants/#1_T_0.2_alpha_0.pdf}
        \caption{$T=0.2$, $\alpha=0$}
    \end{subfigure}
    \hfil
    \begin{subfigure}{0.24\textwidth}
        \centering
        \includegraphics[width=1\linewidth]{figures/cumulants/#1_T_0.2_alpha_pi_over_6.pdf}
        \caption{$T=0.2$, $\alpha=\frac{\pi}{6}$}
    \end{subfigure}
    \hfil
    \begin{subfigure}{0.24\textwidth}
        \centering
        \includegraphics[width=1\linewidth]{figures/cumulants/#1_T_0.2_alpha_pi_over_4.pdf}
        \caption{$T=0.2$, $\alpha=\frac{\pi}{4}$}
    \end{subfigure}
    \hfil
    \begin{subfigure}{0.24\textwidth}
        \centering
        \includegraphics[width=1\linewidth]{figures/cumulants/#1_T_0.2_alpha_pi_over_3.pdf}
        \caption{$T=0.2$, $\alpha=\frac{\pi}{3}$}
    \end{subfigure}

    \begin{subfigure}{0.24\textwidth}
        \centering
        \includegraphics[width=1\linewidth]{figures/cumulants/#1_T_2.0_alpha_0.pdf}
        \caption{$T=2.0$, $\alpha=0$}
    \end{subfigure}
    \hfil
    \begin{subfigure}{0.24\textwidth}
        \centering
        \includegraphics[width=1\linewidth]{figures/cumulants/#1_T_2.0_alpha_pi_over_6.pdf}
        \caption{$T=2.0$, $\alpha=\frac{\pi}{6}$}
    \end{subfigure}
    \hfil
    \begin{subfigure}{0.24\textwidth}
        \centering
        \includegraphics[width=1\linewidth]{figures/cumulants/#1_T_2.0_alpha_pi_over_4.pdf}
        \caption{$T=2.0$, $\alpha=\frac{\pi}{4}$}
    \end{subfigure}
    \hfil
    \begin{subfigure}{0.24\textwidth}
        \centering
        \includegraphics[width=1\linewidth]{figures/cumulants/#1_T_2.0_alpha_pi_over_3.pdf}
        \caption{$T=2.0$, $\alpha=\frac{\pi}{3}$}
    \end{subfigure}

    \begin{subfigure}{0.24\textwidth}
        \centering
        \includegraphics[width=1\linewidth]{figures/cumulants/#1_T_10.0_alpha_0.pdf}
        \caption{$T=10.0$, $\alpha=0$}
    \end{subfigure}
    \hfil
    \begin{subfigure}{0.24\textwidth}
        \centering
        \includegraphics[width=1\linewidth]{figures/cumulants/#1_T_10.0_alpha_pi_over_6.pdf}
        \caption{$T=10.0$, $\alpha=\frac{\pi}{6}$}
    \end{subfigure}
    \hfil
    \begin{subfigure}{0.24\textwidth}
        \centering
        \includegraphics[width=1\linewidth]{figures/cumulants/#1_T_10.0_alpha_pi_over_4.pdf}
        \caption{$T=10.0$, $\alpha=\frac{\pi}{4}$}
    \end{subfigure}
    \hfil
    \begin{subfigure}{0.24\textwidth}
        \centering
        \includegraphics[width=1\linewidth]{figures/cumulants/#1_T_10.0_alpha_pi_over_3.pdf}
        \caption{$T=10.0$, $\alpha=\frac{\pi}{3}$}
    \end{subfigure}
    
    \caption{#2}
    \label{fig:#1}
\end{figure}
}

\begin{document}

\frenchspacing

\title{Full counting statistics in the sine--Gordon model}

\author{Botond C. Nagy}\email{botond.nagy@edu.bme.hu}
\affiliation{Department of Theoretical Physics, Institute of Physics, Budapest University of Technology and Economics, H-1111 Budapest, M{\H u}egyetem rkp.~3.}
\affiliation{
BME-MTA Momentum Statistical Field Theory Research Group, Institute of Physics, Budapest University of Technology and Economics, H-1111 Budapest, M{\H u}egyetem rkp.~3.}
\author{M\'arton Kormos}\email{kormos.marton@ttk.bme.hu}
\affiliation{Department of Theoretical Physics, Institute of Physics, Budapest University of Technology and Economics, H-1111 Budapest, M{\H u}egyetem rkp.~3.}
\affiliation{
BME-MTA Momentum Statistical Field Theory Research Group, Institute of Physics, Budapest University of Technology and Economics, H-1111 Budapest, M{\H u}egyetem rkp.~3.}
\affiliation{HUN-REN-BME-BCE Quantum Technology Research Group, Institute of Physics, Budapest University of Technology and Economics, M{\H u}egyetem rkp.~3., H-1111 Budapest, Hungary}
\author{G\'abor Tak\'acs}\email{takacs.gabor@ttk.bme.hu}
\affiliation{Department of Theoretical Physics, Institute of Physics, Budapest University of Technology and Economics, H-1111 Budapest, M{\H u}egyetem rkp.~3.}
\affiliation{
BME-MTA Momentum Statistical Field Theory Research Group, Institute of Physics, Budapest University of Technology and Economics, H-1111 Budapest, M{\H u}egyetem rkp.~3.}

\date{July 23, 2026}

\begin{abstract} 
Full counting statistics (FCS) is a dynamical generalisation of the free energy, encapsulating detailed information about the distribution and large-scale correlation functions of conserved charges and their associated currents. 
In this work, we present a comprehensive numerical study of the FCS and the cumulants of the three lowest charges across the full parameter space of the sine--Gordon field theory.
To this end, we extend the thermodynamic Bethe Ansatz (TBA) formulation of the FCS to the sine--Gordon model, emphasise the methodological subtleties for a reliable numerical implementation, and compare numerical results with analytical predictions in certain limits.
\end{abstract} 

\maketitle 
\tableofcontents

\section{Introduction}
Characterising correlations and fluctuations in interacting quantum systems is a notoriously difficult task and has been the subject of considerable theoretical and experimental effort in recent decades. The simplest observables, such as expectation values and two-point correlation functions, probe the system at the level of linear response. However, much more information is encoded in the higher-point correlations and in the full distribution functions of integrated observables. The latter often obeys a large deviation principle that describes rare but large fluctuations \cite{Touchette2009}. In order to gain information about the dynamics of the system and the properties of quasiparticle excitations, one should consider dynamical quantities such as the probability distribution of the time-integrated current of some conserved quantity, also known as the full counting statistics (FCS) \cite{Esposito2009}. 

Recently, there has been growing interest in FCS and related quantities due to their increasing experimental accessibility, both in cold-atom systems \cite{Bouchoule2006,Bouchoule2010,Gritsev2007a,Schweigler2017,Wei2022} and in quantum computers \cite{Fan2024,Rosenberg2024,Samajdar2024}. Analytical results for the FCS have been obtained in quadratic systems \cite{Levitov1993,Doyon2015,Yoshimura2018} and in conformal field theories \cite{Bernard2016}. It has also been investigated in interacting quantum systems, both integrable \cite{Lamacraft2008,Stephan2017,Collura2017,Groha2018,Bastianello2018,GangardtFCS2019,Perfetto2019,Ares2021,Krajnik2024,Valli2025} and chaotic \cite{McCulloch2023,Gopalakrishnan2024}, and it has been shown that it can be used to define dynamical universality classes \cite{Gopalakrishnan2024,Krajnik2024}. A major advance in the theoretical understanding of the FCS in integrable systems was the development of the so-called Ballistic Fluctuation Theory \cite{Myers2020,Doyon2019b}. It applies to integrable systems with ballistic transport and provides an analytic framework for describing fluctuations on large, hydrodynamic scales. 

In this work, we explore fluctuations in the sine--Gordon field theory via the full distribution functions of space-time integrals of conserved quantities and their currents. The integrable sine--Gordon model has a wide range of applications as the low-energy effective description of a broad spectrum of gapped one-dimensional systems \cite{Giamarchi:743140,Essler2005} such as carbon nanotubes, quasi-1D antiferromagnets \cite{Zvyagin2004} or organic conductors \cite{Controzzi2001}, trapped ultra-cold atoms \cite{Gritsev2007,Cirac2010,Haller2010,Wybo2023}, coupled spin chains \cite{Wybo2022} and quantum circuits \cite{Roy2021}. Importantly, the model can be realised in tunnel-coupled quasi-1D condensates of ultracold atoms \cite{Schweigler2017,Zache2020,2024PhRvB.109c5118B}, where the phase fluctuations can be studied via matter-wave interferometry, which is the main motivation for our work.

The transport of energy, momentum, and topological charge is ballistic, as signalled by the non-zero Drude weights \cite{2023PhRvB.108x1105N,2024ScPP...16..145N}. Our strategy is to apply the BFT theory, taking advantage of the recently obtained thermodynamic and hydrodynamic description\footnote{There is an alternative description of the equilibrium thermodynamics of the model in terms of a nonlinear integral equations \cite{1987NuPhB.290..363D,
1995NuPhB.438..413D,Hegedus2025,Hegedus2026}, however, it is not known how to formulate hydrodynamics in this framework.} of the sine--Gordon model for generic couplings \cite{2023PhRvB.108x1105N,2024ScPP...16..145N} to provide a comprehensive study of the fluctuations of various conserved quantities and their currents. Our work continues and extends the analysis of Ref. \cite{DelVecchio2023} carried out in the weak-coupling semiclassical regime of the theory based on the hydrodynamic description of the classical sine--Gordon model \cite{koch2023exact}. One of the main results of our work is that, similarly to its Drude weight, the topological charge's cumulants depend in a very irregular, fractal-like manner on the coupling strength of the model.

The paper is organised as follows. Section \ref{sec:sG_TBA} provides an overview of the key features of the sine--Gordon model and summarises the thermodynamic Bethe Ansatz description of the theory, which provides the numerical framework for our investigations. Section \ref{sec:FCS} introduces the primary object of our study, the full counting statistics, and presents the TBA expressions for its evaluation. We present and discuss our results for the cumulants and the full counting statistics in section \ref{sec:results}. 
Additional plots illustrating the behaviour of quantities that receive less detailed discussion in the main text, together with supplementary derivations, are provided in the appendix.

\section{The sine--Gordon model and its thermodynamic description}
\label{sec:sG_TBA}
The sine--Gordon model is defined via the following Hamiltonian
\begin{equation}
    H = \int \text{d}x \left[\frac{1}{2}\left(\partial_t \phi\right)^2 + \frac{1}{2}\left(\partial_x \phi\right)^2 - \lambda \cos(\beta\phi)\right]\,.
\end{equation}
In this equation, $\lambda$ sets the energy scale of the theory and $\beta$ is the coupling strength. It is customary to introduce the so-called renormalised coupling $\xi=\frac{\beta^2}{8\pi-\beta^2}$, in terms of which some properties of the theory can be expressed more easily.
The sine--Gordon model is an integrable field theory, meaning that it possesses an infinite number of conserved charges. In this work, we consider only the energy, momentum, and topological charge. They are defined as
\begin{subequations}
\begin{alignat}{3}
    h^e&=\frac{1}{2}\left(\partial_t \phi\right)^2 + \frac{1}{2}\left(\partial_x \phi\right)^2 - \lambda \cos(\beta\phi)\,, \qquad
    &&j^e=-\left(\partial_t \phi\right)\left(\partial_x \phi\right)\,, \\
    h^p&=-\left(\partial_t \phi\right)\left(\partial_x \phi\right)\,,\qquad 
    &&j^p=\frac{1}{2}\left(\partial_t \phi\right)^2 + \frac{1}{2}\left(\partial_x \phi\right)^2 + \lambda \cos(\beta\phi)\,, \\
    h^q&=\frac{\beta}{2\pi}\partial_x\phi\,,\qquad &&j^q=-\frac{\beta}{2\pi}\partial_t\phi\,. \label{eq:topqj}
\end{alignat}
\end{subequations}
Energy and momentum are conserved by spacetime translation symmetry, while the topological charge is conserved by the symmetry of second derivatives.

For any value of the coupling, the theory contains two elementary massive excitations: the soliton ($S$) and the antisoliton ($\overline{S}$), which have the same mass $m_S$ but carry opposite topological charges.
Based on the value of $\beta$, the theory exhibits two regimes and may contain additional stable particles. For $\beta^2/8\pi>1/2$, the $S-\overline{S}$ interaction is repulsive, and the only particles in the spectrum are the soliton and the antisoliton.
For $\beta^2/8\pi < 1/2$, the $S-\overline{S}$ interaction is attractive, and the spectrum also contains bound states $B_k$, known as breathers. 
The number of different breathers is $n_B=\lceil1/\xi-1\rceil$\footnote{$\lceil.\rceil$ denotes the ceiling function}, and their masses are $m_{B_k}=2m_S\sin(k\pi\xi/2)$, in terms of the soliton mass $m_S$. For a more detailed description of the model, see e.g. Ref. \cite{2024ScPP...16..145N}.

Due to the model's integrability, its scattering matrix is factorised. 
For generic couplings, SS scattering is non-diagonal, which can be resolved by using the nested Bethe Ansatz. The idea is to diagonalise the transfer matrix, which requires introducing auxiliary particles called elementary magnons. These act as ``spin-flips" in topological charge, i.e. they transform solitons to antisolitons, and as such, eliminate the need for antisolitons in the description, while in exchange, they carry topological charge but no energy and momentum. Bound states of elementary magnons are referred to as magnons, and the number of different types of magnon bound states depends on the renormalised coupling in the following way. Writing the coupling as a continued fraction,
\begin{equation}
    \xi = \frac{\beta^2}{8\pi-\beta^2} = \frac{1}{\displaystyle n_B + \frac{1}{\displaystyle \nu_1 + \frac{1}{\displaystyle \nu_2+\dots}}}\,.
\end{equation}
the TBA spectrum contains 1 soliton, $n_B$ breathers, and $\sum_i\nu_i$ magnons, where $i$ is referred to as a level index. The properties of the magnons depend on $\xi$ in an intricate way, and the full description can be found in \cite{2024ScPP...16..145N}.

The equilibrium states are characterised by the filling fraction for each particle species $a$, defined as the ratio of the occupied and the total density of states
\begin{equation}
    \vartheta_a(\theta) = \frac{\rho_a(\theta)}{\rho_a^{\text{tot}}(\theta)} = \frac{1}{1+e^{\epsilon_a(\theta)}}\,.
    \label{eq:filling}
\end{equation}
The second equation defines the pseudo-energies, which obey the following coupled set of integral equations: 
\begin{equation}
    \epsilon_a = w_a - \sum_b \eta_b \Phi_{ab} * \log\left(1+e^{-\epsilon_b}\right)\,.
    \label{eq:tba_pseudo-energy}
\end{equation}
In a generalised Gibbs ensemble, the source terms are $w_a=\sum_j\beta_j \,h_a^{(j)}(\theta)$, where  $\beta_j$ -- not to be confused with the coupling strength $\beta$ which doesn't have an index -- is the generalised chemical potential (Lagrange multiplier) associated with the $j$th conserved charge whose eigenvalues on the single particle states of particle $a$ are $h^{(j)}_a(\theta)$. 
In the sine--Gordon model, these are known explicitly for all conserved charges \cite{Hegedus2026}. In this work, we focus on the momentum, $h^p_a\equiv p_a = m_a\sinh\theta$, the energy, $h^e_a\equiv e_a = m_a\cosh\theta$ with $\beta_e=1/T$, and the topological charge $h^q_a\equiv q_a$ with $\beta_q=-\mu/T$, where $\mu$ is the chemical potential.
The signs $\eta_a=\pm 1$ are $+1$ for massive particles, but are non-trivial for the magnons. The convolution kernels $\Phi_{ab}$ depend on the scattering matrices of the theory. Detailed expressions of these quantities can be found in \cite{2024ScPP...16..145N}. The (generalised) free energy density (per temperature) is given by\footnote{Note that while in principle Eq. \eqref{eq:TBA_free_energy} should contain the mass $m_a$ multiplied by the sign factor $\eta_a$, the latter can be omitted as $\eta_a=+1$ for all massive nodes.}
\begin{equation}
    \tilde f \equiv \frac{f}{T} = -\frac1L\log Z= -\sum_a \int \frac{\text{d}\theta}{2\pi}m_a\cosh\theta\log\left(1+e^{-\epsilon_a}\right)\,.
\label{eq:TBA_free_energy}\end{equation}
Once the filling is determined from Eqs.(\ref{eq:filling},\ref{eq:tba_pseudo-energy}), the dressed charges are calculated as
\begin{equation}
    h_a^{(j)\text{dr}} = \eta_a\frac{\partial \epsilon_a}{\partial \beta_j}=\eta_a h_a^{(j)} + \eta_a \sum_b \Phi_{ab}*\vartheta_b h_b^{(j)\text{dr}}\,.
    \label{eq:tba_dressing}
\end{equation}
The dressed charges are instrumental in calculating the thermodynamic expectation values of the corresponding observables.
As a special case of the dressing equation, the total density of states is \begin{equation}
    \rho_a^{\text{tot}}(\theta) = \eta_a\frac{p_a'^{\text{dr}}(\theta)}{2\pi}\,,
\label{eq:particle_densities}
\end{equation}
and $\rho_a(\theta)=\rho^{\text{tot}}_a(\theta)\vartheta_a(\theta)$ is the density of (occupied) states.
We note that it is possible to partially decouple \cite{2024ScPP...16..145N} the above sets of equations (\ref{eq:tba_pseudo-energy},\ref{eq:tba_dressing}) for numerical efficiency. However, the decoupling procedure mixes the source terms in Eqs. (\ref{eq:tba_pseudo-energy}) and (\ref{eq:tba_dressing}), and for complicated quantities to be dressed, the partially decoupled version of the equations becomes cumbersome to use. For this reason, we used the fully coupled version in this work.

The expectation values of the conserved charge densities are generated by the (generalised) free energy: $\langle \hat h^{(k)}\rangle = \partial_{\beta_k} \tilde f$. In particular, in the grand canonical ensemble, the energy and topological charge density are given by
\begin{equation}
    \langle e\rangle = \frac{\partial \tilde f(R,\bar\mu)}{\partial R}\,,\qquad \langle q\rangle = -\frac{\partial \tilde f(R,\bar\mu)}{\partial \bar\mu}\,,
    \label{eq:fTderiv}
\end{equation}
where $R=1/T$ and $\bar\mu=\mu/T$. Similarly, the current expectation values are generated by the so-called free energy flux, $\langle \hat j^{(k)}\rangle = \partial_{\beta_k} \tilde g$, where \cite{2016PhRvX...6d1065C}
\begin{equation}
    \tilde g = -\sum_a \int \frac{\text{d}\theta}{2\pi}\eta_a m_a\sinh\theta\log\left(1+e^{-\epsilon_a}\right)\,.
\label{eq:flux}
\end{equation}
The expectation values of the charges and their currents can be written explicitly as (suppressing the label of the charge)
\begin{subequations}
\begin{align}
\begin{split}
    \langle h\rangle &= \sum_a\int \text{d}\theta\ \rho_a(\theta)h_a(\theta) = \sum_a \int \text{d}\theta\ \frac{p_a'(\theta)}{2\pi}\vartheta_a(\theta)h_a^{\text{dr}}(\theta)\,,
    \label{eq:tba_charge_eq}
\end{split}\\
\begin{split}
    \langle j \rangle &= \sum_a \int \text{d}\theta\ \rho_a(\theta)h_a(\theta)v_a^{\text{eff}}(\theta) = 
    \sum_a \int \text{d}\theta\ \frac{e_a'(\theta)}{2\pi}\vartheta_a(\theta)h_a^{\text{dr}}(\theta)\,.
    \label{eq:tba_current_eq}
\end{split}
\end{align}
\end{subequations}
The expression for current densities was originally conjectured in \cite{2016PhRvX...6d1065C,2016PhRvL.117t7201B} and proved later in \cite{2020PhRvX..10a1054B}.
In the expression of the current, we introduced the effective velocity, which is calculated as
\begin{equation}
    v_a^{\text{eff}}(\theta) = \frac{e_a'^{\text{dr}}(\theta)}{p_a'^{\text{dr}}(\theta)}\,.
\end{equation}
The effective velocity accounts for the fact that particles propagate through a finite-density background, and interactions with the sea of other particles modify their large-scale velocity.

\section{Full counting statistics}
\label{sec:FCS}
Let us now turn to the main objective of our work: describing the fluctuations of the conserved densities and their currents through their full distribution functions and their cumulants. The path towards these quantities is provided by the Ballistic Fluctuation Theory (BFT) \cite{Myers2020,Doyon2019b}, which captures the fluctuations of conserved quantities obeying ballistic transport. In our earlier works \cite{2023PhRvB.108x1105N,2024ScPP...16..145N}, we demonstrated that for rational couplings, the Drude weights for energy, momentum, and topological charge are finite, indicating ballistic transport, and thus the conditions for applying the BFT framework are fulfilled. 

In the following, we briefly summarise the BFT approach. It is convenient to combine the integrated charge density and current density of the $i$th charge into a more general quantity defined as a line integral in space-time along a curve $(x(s),t(s))$ parametrised by $s\in[0,1]$:
\begin{equation}
    J^{(i)}_\gamma  
    = \int_{s=0}^{1} \text{d} s \left(\dot t(s)\, j^{(i)}[x(s),t(s)] -  \dot x(s)\, h^{(i)}[x(s),t(s)]\right)
    \,.
    \label{eq:J}
\end{equation}
Due to the continuity equation, this quantity is path-independent and only depends on the endpoints. If it is calculated in a space-time translationally invariant state, then it only depends on the space-time separation that we can parametrise by $x(s)-x(0)=\ell\sin\alpha$, $t(s)-t(0)=\ell\cos\alpha$, such that the angle $\alpha$, determining the space-time direction, is measured from the temporal axis. 
In particular, for $\alpha=0$ and $\alpha=\pi/2$ we recover the integrated current (transported charge) over time $\ell$ and the charge contained in an interval of length $\ell$, respectively (note that the speed of light is set to $c=1$):
\begin{equation}
\begin{split}
    J^{(i)}(\ell,\alpha=0)=\int_{-\ell/2}^{\ell/2}\text{d}t\ j^{(i)}(0,t)\,,\qquad
    J^{(i)}(\ell,\alpha=\pi/2)=-\int_{-\ell/2}^{\ell/2}\text{d}x\ h^{(i)}(x,0)\,,
\end{split}
\end{equation}

The full probability distribution of this quantity can be described by its characteristic (moment generating) function $\Big\langle \exp\Big[\lambda J^{(i)}(\ell,\alpha) \Big]  \Big\rangle$ where $\lambda$ is a counting variable. It is the Laplace transform of the probability distribution (or Fourier transform for imaginary $\lambda$). The moments of the distribution can be obtained by taking derivatives with respect to $\lambda$ at $\lambda=0$, while the cumulants can be generated by taking derivatives of its logarithm.

The BFT predicts that for large $\ell$, the characteristic function has the asymptotic behaviour 
\begin{equation}
    \Big\langle \exp\Big[\lambda J^{(i)}(\ell,\alpha) \Big]  \Big\rangle \asymp \exp\big[\ell F^{(i)}_{\alpha}(\lambda)\big]\,.
    \label{eq:charfnasymp}
\end{equation}
That is, all the cumulants scale extensively in $\ell$, which incorporates the same scaling in $x$ and $t$ in accordance with the ballistic dynamics. Under certain analyticity conditions \cite{Krajnik2022,Krajnik2024,Gopalakrishnan2024} which we assume to hold, the cumulant densities (scaled cumulants) are generated by the ``dynamical free energy'' $F^{(i)}_{\alpha}(\lambda)$ as
\begin{equation}
    c^{(i)}_n(\alpha) = \frac{\text{d}^nF^{(i)}_{\alpha}(\lambda)}{\text{d}\lambda^n} \Bigg\rvert_{\lambda=0}\,.
    \label{eq:cn}
\end{equation}
The quantity $F^{(i)}_{\alpha}(\lambda)$, which we shall refer to as the Full Counting Statistics (FCS), is a function of $\lambda$ and only depends on the space-time direction, and it is the central object of the theory.
The scaled cumulants \eqref{eq:cn} carry information about the connected multi-point correlation functions, for example
\begin{subequations}
\begin{align}
\begin{split}
    c_n^{(i)}(0) &= \left\langle \int_{-\ell/2}^{\ell/2} \text{d}t_1\dots\text{d}t_{n-1}\ j^{(i)}(0,0)j^{(i)}(0,t_1)\dots j^{(i)}(0,t_{n-1}) \right\rangle_\text{conn},
\end{split}\\
\begin{split}
    c_n^{(i)}
    (\pi/2)
    &= \left\langle \int_{-\ell/2}^{\ell/2}\text{d}x_1\dots\text{d}x_{n-1}\ h^{(i)}(0,0)h^{(i)}(x_1,0)\dots h^{(i)}(x_{n-1},0) \right\rangle_\text{conn}.
\end{split}
\end{align}
\end{subequations}

Before discussing how to calculate the FCS, we make a brief digression on the special case of the topological charge in the sine--Gordon model. Due to the relations \eqref{eq:topqj} it follows that 
\begin{equation}
    \int_{(0,0)}^{(x,t)}\Big[j^q(x',t')\text{d}t'-h^q(x',t')\text{d}x'\Big] = -\frac{\beta}{2\pi}\big[\phi(x,t)-\phi(0,0)\big]\,,
\end{equation}
i.e. the random variable \eqref{eq:J} is simply proportional to the difference of field values at the endpoints of the interval \cite{Doyon2019b,DelVecchio2023}. 
Then the characteristic function becomes
\begin{equation}
    \Big\langle \exp\Big[\lambda J^q(\ell,\alpha) \Big]  \Big\rangle
    =\left\langle e^{-\lambda\beta/2\pi [\phi(x,t)-\phi(0,0)]}\right\rangle\,.
\end{equation}
For space-like separations $|x|>|t|$, the field operators in the exponent commute, and we can split the exponential into the product of two exponentials of the field. One can argue that even for time-like separations, the corrections coming from the Baker--Campbell--Hausdorff formula yield power-law corrections and do not affect the leading exponential decay \cite{2022JSMTE2022e3102D}. Analytically continuing to imaginary values $\lambda=-i\nu$, we arrive at the two-point correlation function of vertex operators $e^{\pm i\nu\beta\phi/2\pi}$ in a generalised Gibbs state. BFT thus predicts exponential decay at large separations and provides access to the inverse correlation length $F^q_\alpha(-i\nu)$.

Returning to the general BFT theory, the FCS is given by an integral of current and charge expectation values along a path in the space of generalised Gibbs ensembles (GGE-s):
\begin{equation}
    F^{(i)}_\alpha(\lambda) = \int_0 ^\lambda \text{d}\lambda' \left(\cos\alpha\langle j^{(i)} \rangle_{\lambda'}-\sin\alpha\langle h^{(i)}\rangle_{\lambda'}\right)\,,
    \label{eq:FCS}
\end{equation}
where $\lambda'$ parametrises the states along the path. The averages $\langle \dots\rangle_{\lambda'}$ appearing in Eq.~\eqref{eq:FCS} are taken in these states that are characterised by their $\lambda'$-dependent pseudo-energies $\epsilon_a^{(i)}(\theta,\lambda')$. The path in the space of GGE-s is determined by the flow equations for the pseudo-energies
\begin{equation}
\begin{split}
    \partial_{\lambda'} \epsilon_a^{(i)}(\theta,\lambda') &= \text{sign}\left(\sin\alpha - v_a^{\text{eff}}(\theta,\lambda)\cos\alpha\right){h^{(i)\text{dr}}_a}(\theta,\lambda')\,, \\
    \epsilon_a^{(i)}(\theta,0) &= \epsilon_a(\theta)\,,
\end{split}
\label{eq:flow_eq}
\end{equation}
where $v_a^{\text{eff}}$ and ${h^{(i)\text{dr}}_a}$ also acquire a $\lambda'$-dependence, since they are computed in the deformed state corresponding to the $\lambda'$-dependent pseudo-energies.
Once the pseudo-energies are determined, the charge and current expectation values are calculated in the same way as in (\ref{eq:tba_charge_eq},\ref{eq:tba_current_eq}).

The flow equation can be solved explicitly in those cases where the sign function becomes independent of $\lambda'$. Importantly, this happens for space-like separations, $|x|>t$ ($\alpha>\pi/4$), where the sign simply agrees with the sign of $x$. Then due to the relation ${h^{(j)\text{dr}}_a} = \eta_a\partial \epsilon_a/\partial \beta_j$ (where $\eta_a=\pm1$), the solution amounts to a shift in the $\beta_j$ chemical potentials. The charges (currents) are obtained by taking derivatives of the free energy \eqref{eq:TBA_free_energy} (free energy flux \eqref{eq:flux}) with respect to the $\beta_j$ variables, $\langle \hat h^{(k)}\rangle = \partial_{\beta_k} \tilde f$, $\langle \hat j^{(k)}\rangle = \partial_{\beta_k} \tilde g$. This is now equivalent to taking derivatives with respect to $\lambda$. Consequently, the integral over $\lambda'$ in Eq. \eqref{eq:FCS} simply yields the difference of the free energy and free energy flux at chemical potentials $\beta_j\pm\lambda$ and $\beta_j$.

Returning to the general case, it is possible to derive explicit expressions for the scaled cumulants \eqref{eq:cn}. In terms of the shorthand notations\footnote{In the following, we omit the index $i$ of the charge when the charge is clear from the context.}
\begin{equation}
    H:=h^{\text{dr}}\,,\quad s = \text{sign}\left(\cos\alpha\ v^{\text{eff}}-\sin\alpha \right)\,,\qquad
    \tilde{f}=-\left(\frac{1}{f}\frac{\partial f}{\partial \epsilon} + 2f\right)\,, \qquad
    \hat{f} = -\left(\frac{1}{f\tilde{f}}\frac{\partial \left(f\tilde{f}\right)}{\partial \epsilon} + 3f\right)\,,
\end{equation}
the expressions for the first four cumulants in terms of TBA quantities read (see also Appendix \ref{sec:app_cumulants_tba})
\begin{subequations}
\label{eq:main_cs}
\begin{align}
    c_1&= \int_a \frac{\text{d}\theta}{2\pi} \left(\cos\alpha\ e' -\sin\alpha\ p'\right)\vartheta H\,,\quad e=m\cosh\theta\,,\qquad p=m\sinh\theta
    \label{eq:main_c1}
\\
    c_2 &= \int_a \text{d}\theta\ \rho f \left|\cos\alpha\ v^{\text{eff}} -\sin\alpha\right| H^2\,,
    \label{eq:main_c2}
\\
    c_3 &= \int_a \text{d}\theta\ \rho f \left|\cos\alpha\ v^{\text{eff}} -\sin\alpha\right|H\left[ 3\left(sfH^2\right)^{\text{dr}}+\eta s \tilde{f}H^2 \right]\,,
    \label{eq:main_c3}
\\
    c_4 &= \int_a \text{d}\theta\ \rho f \left|\cos\alpha\ v^{\text{eff}} -\sin\alpha\right|\times\nonumber
    \\ 
    &\times \bigg\{
    \tilde{f}\hat{f}H^4 +
    3 \left[\left(sfH^2\right)^{\text{dr}}\right]^2 +
    4 H \left( \eta f \tilde{f} H^3 \right)^{\text{dr}} + 
    6 \eta s \tilde{f} H^2 \left(sfH^2\right)^{\text{dr}} + 
    12 H \left[sfH\left(sfH^2\right)^{\text{dr}}\right]^{\text{dr}}
    \bigg\}\,,\label{eq:main_c4}
\end{align}
\end{subequations}
where we simplified the notation by suppressing the particle indices. As a result, the integrations must be understood as containing an implicit summation over the particle index, $\int_a \equiv \sum_a \int$. The expressions for higher cumulants become increasingly involved. Note that these equations were originally derived in \cite{Myers2020}, but in the case of the sine--Gordon model, they are supplemented by additional sign factors $\eta$. The consistent incorporation of these signs represents an essential generalisation required for the correct treatment of the sine--Gordon model.

\section{Results}
\label{sec:results}

In this section, we present our numerical results for the temperature, coupling strength, and $\alpha$-dependence of the first four cumulants associated with the topological charge, momentum, and energy in canonical equilibrium states.

As noted in Section \ref{sec:sG_TBA}, there exists a partially decoupled formulation of the TBA equations \cite{2024ScPP...16..145N}, which offers improved numerical performance. However, the decoupling procedure entails mixing the source terms in Eq. (\ref{eq:tba_dressing}). In \cite{2024ScPP...16..145N}, this mixing was followed only for the topological charge, momentum and energy. When dressing more complicated quantities, such as those relevant to cumulants, employing the fully coupled dressing equations is the more straightforward and reliable approach. Although the transformation of the source terms can, in principle, be determined by following the established procedure, partial decoupling yields a meaningful simplification only for relatively simple source terms. By contrast, the quantities appearing in the cumulant expressions are already highly intricate before dressing, and several terms require these quantities to be dressed multiple times. Consequently, the resulting partially decoupled expressions become considerably more cumbersome, outweighing any potential simplification and rendering this approach impractical.

One possible strategy for computing the cumulants is to evaluate the FCS from Eq. (\ref{eq:FCS}) and subsequently determine the cumulants by numerical differentiation. This method was proposed in \cite{Myers2020, DelVecchio2023}. However, our analysis indicates that only the first derivative can be obtained with satisfactory numerical precision. For instance, while $c_2$ may be computed directly from Eq.(\ref{eq:main_c2}) or by keeping the $\lambda$-dependence in Eq.(\ref{eq:main_c1}) and taking a single numerical derivative at $\lambda=0$, attempting to extract $c_2$ by taking a second numerical derivative of Eq.(\ref{eq:FCS}) leads to noticeably reduced accuracy. This loss of precision becomes progressively more severe for higher-order derivatives, see Appendix \ref{app:comparison_subsec} for more details.

For an extensive collection of plots illustrating the functional dependence of the cumulants on temperature, coupling strength, and space-time direction, we refer the interested reader to Appendix \ref{sec:gen_obs}.
In the plots and tables, dimensionful quantities are measured in appropriate powers of the soliton mass (e.g. ``$T=0.1$'' means $T=0.1m_S$ etc.).

\subsection{Analytically computable limits}

In this section, we present the comparison of the numerical evaluation of the cumulants with analytical predictions in analytically tractable parameter regimes.

\subsubsection{Reflectionless points in the low temperature limit}
\label{sec:refless_lowT}

For integer values of $1/\xi$, the scattering of solitons is diagonal and the TBA equations \eqref{eq:tba_pseudo-energy} can be written in terms of the massive soliton, antisoliton and breather particles, and there is no need to introduce magnons. At very low temperatures, the source terms dominate, and the convolution terms can be neglected, resulting in free-fermionic filling fractions and root densities. Dressed quantities are equal to their bare values, e.g. $h_a^{q\,\text{dr}}=\pm 1$ for the soliton and antisoliton, and zero for the breathers. The flow equations \eqref{eq:flow_eq} become trivial: the pseudo-energies are linear functions of $\lambda$. Plugging in the charge and current expression from Eqs. \eqref{eq:tba_charge_eq}, \eqref{eq:tba_current_eq} to the FCS formula \eqref{eq:FCS}, the integral over $\lambda$ can be computed, which leads to a single rapidity integral. In the case of the topological charge, the result is
\begin{multline}
    F^q(\lambda) = \int \frac{\text{d}\theta}{2\pi}m_S\cosh\theta |\sin\alpha-\cos\alpha\tanh\theta| \times \\
    \times \left[\log\left(1+e^{-m_S\cosh\theta/T+\lambda s}\right)+\log\left(1+e^{-m_S\cosh\theta/T-\lambda s}\right)-2\log\left(1+e^{-m_S\cosh\theta/T}\right)\right]\,,
\end{multline}
where $s=\mathrm{sign}(\cos\alpha\tanh\theta-\sin\alpha)$. Since the expression is symmetric in $s=\pm1$, we can replace $s$ by 1. Expanding the integrand to the leading order in $e^{-m_S/T}$, we arrive at
\begin{equation}
    \ell F^q(\lambda) = 4\sinh^2(\lambda/2) \int \frac{\text{d}\theta}{2\pi}m_S\cosh\theta e^{-m_S\cosh\theta/T} |x-v(\theta)t| \equiv 2\sinh^2(\lambda/2)\, \Omega_{x,t}\,,
    \label{eq:lowT}
\end{equation}
where $v(\theta)=\tanh\theta$. This result has a nice semiclassical interpretation. Taking into account that at very low temperatures, the soliton and antisoliton densities approach the Maxwell--Boltzmann distribution\footnote{Note that the number of solitons and antisolitons is not separately fixed (there is no chemical potential for them) and at low temperatures their density is low, which is the hallmark of a semiclassical regime.}, the integral $\Omega_{x,t}$ counts the mean number of soliton and antisoliton trajectories, assumed to be straight lines, that cross the segment connecting the space-time points $(0,0)$ and $(x,t)=\ell(\sin\alpha,\cos\alpha)$. Indeed, the result \eqref{eq:lowT} agrees with the semiclassical result for the vertex operator two-point function in the case of reflectionless scattering \cite{2005PhRvL..95r7201D,PhysRevB.106.205151}.

\begin{figure}[t]
    \centering
    \includegraphics[height=3.5cm]    {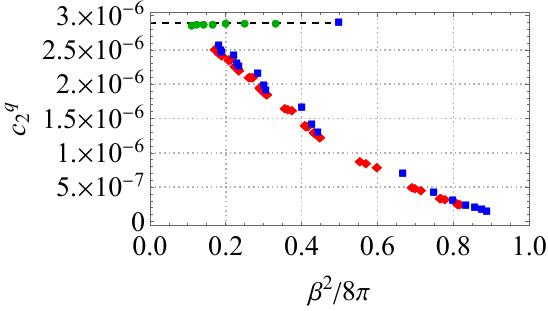}
    \hfil
    \includegraphics[height=3.5cm] {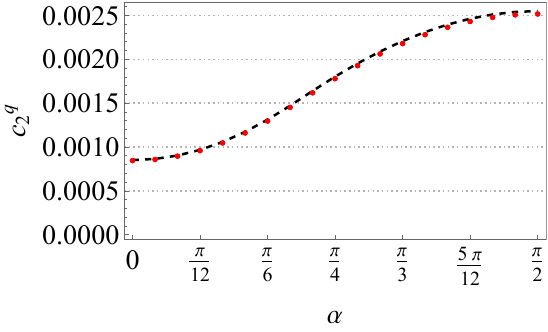}
    \caption{Variance $c_2^q$ as a function of the coupling at $T=0.1$, $\alpha=0$ (left) and as a function of $\alpha$ at $T=0.2$, $\xi=1/3$ (right). The black dashed lines are the second cumulant derived from Eq.(\ref{eq:lowT}). In the left panel, the colours refer to the number of magnonic levels.}
    \label{fig:low_T_reflectionless}
\end{figure}

In Fig. \ref{fig:low_T_reflectionless}, we compare the second cumulant from Eq. \eqref{eq:lowT} with numerical results at $T/m_S=0.1$ and $0.2$. The good agreement demonstrates the validity of the approximate formula \eqref{eq:lowT}.

The $\lambda$-dependence of the result is simple enough to enable the analytical calculation of the distribution function itself. The result is the Skellam distribution,
\begin{equation}
    P[J^{q}(\ell,\alpha)] = e^{-\Omega_{x,t}}I_{|J^q|}(\Omega_{x,t})\,,
\end{equation}
where 
$I_n(u)$ is the modified Bessel function. All the cumulants of the distribution are equal, $\ell c_n = 2\Omega_{x,t}$.

Note that for space-like separations, $x > t$, the absolute value sign can be omitted. Since $v(\theta)$ is an odd function, its integral gives zero, so 
\begin{equation}
    \Omega_{x,t}=2\int \frac{\text{d}\theta}{2\pi}m_S\cosh\theta e^{-m_S\cosh\theta/T} x = n_\text{K}x\,,
\end{equation}
where $n_\text{K}$ is the total density of the kinks (solitons and antisolitons) in the low-temperature limit.

\subsubsection{Level-one repulsive couplings at low temperatures}
\label{sec:level_one_repulsive}

\begin{figure}[H]
    \centering
    \includegraphics[width=0.35\linewidth]{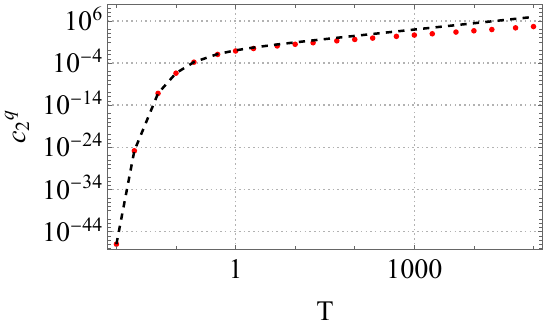}
    \caption{Variance $c_2^q$ for $\alpha=0$, $\xi=3$ as a function of temperature. Note that the $y$ scale is logarithmic and rather wide, so Eq.(\ref{eq:low_T_repulsive_approx}) is only a good approximation at low and medium temperatures. See also Table \ref{table:low_T_repulsive_approx}.}
    \label{fig:low_T_repulsive_alpha_0}
\end{figure}

For $\alpha=0$, at low temperatures and integer $\xi$ couplings when there is a single magnonic level, we can compute the variance of the current (see Appendix \ref{app:c2})
\begin{equation}
    c_2^q(\alpha=0) = \sqrt{\frac{8}{m_S}}\left(\frac{T}{\pi}\right)^{3/2}e^{-m_S/T}\tan\left(\frac{\pi}{2\xi}\right)\,,
    \label{eq:low_T_repulsive_approx}
\end{equation}
which is compared with the numerical results in Table \ref{table:low_T_repulsive_approx} and in Fig. \ref{fig:low_T_repulsive_alpha_0}.
\begin{table}[H]
\centering
\begin{tabular}{|c|c|c|c|c|c|c|c|}
\hline
$T$         & $0.01$                 & $0.02$                 & $0.05$                 & $0.1$                 & $0.2$       & $0.5$      & $1$        \\ \hline
approx.   & $1.088\cdot10^{-47}$   & $1.59124\cdot10^{-25}$ & $6.66914\cdot10^{-12}$ & $4.10313\cdot10^{-7}$ & 0.000167624 & 0.011745   & 0.0761216  \\ \hline
numeric   & $1.09097\cdot10^{-47}$ & $1.59985\cdot10^{-25}$ & $6.7581\cdot10^{-12}$  & $4.21032\cdot10^{-7}$ & 0.000176739 & 0.0140322  & 0.107886   \\ \hline
rel. err. & $0.00272446$           & $0.00541598$           & $0.0133404$            & $0.026123$            & $0.0543764$ & $0.194735$ & $0.417285$ \\ \hline
\end{tabular}
\caption{Comparison of the numerical results for $c_2^q$ for $\alpha=0$, $\xi=3$ with Eq.(\ref{eq:low_T_repulsive_approx}). See also Fig. \ref{fig:low_T_repulsive_alpha_0}.}
\label{table:low_T_repulsive_approx}
\end{table}

\begin{figure}[H]
    \centering

    \begin{subfigure}{0.23\textwidth}
        \includegraphics[width=1\linewidth]{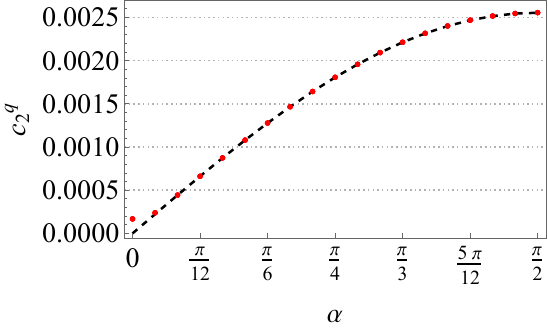}
        \caption{$c_2^q$, $T=0.2$, $\xi=3$}
        \label{fig:low_T_repulsive_alpha_non0_c2q}
    \end{subfigure}
    \hfil
    \begin{subfigure}{0.23\textwidth}
        \includegraphics[width=1\linewidth]{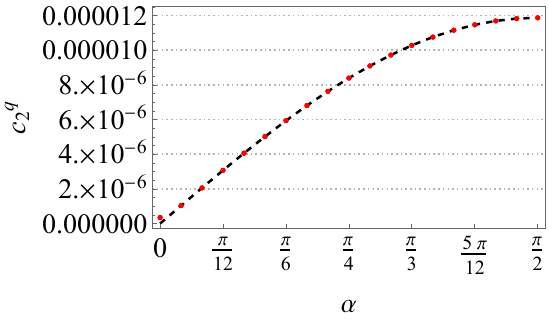}
        \caption{$c_2^q$, $T=0.1$, $\xi=3+1/4$}
        \label{fig:low_T_repulsive_alpha_non0_c2q_level2}
    \end{subfigure}
    \hfil
    \begin{subfigure}{0.23\textwidth}
        \includegraphics[width=1\linewidth]{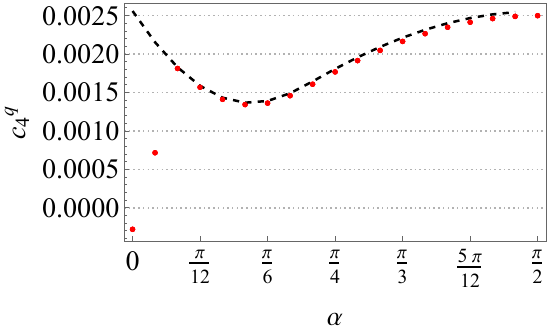}
        \caption{$c_4^q$, $T=0.2$, $\xi=3$}
        \label{fig:low_T_repulsive_alpha_non0_c4q}
    \end{subfigure}
    \hfil
    \begin{subfigure}{0.23\textwidth}
        \includegraphics[width=1\linewidth]{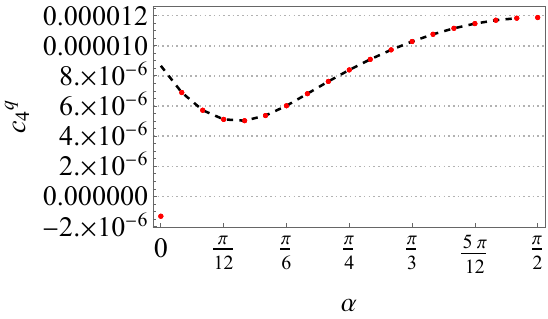}
        \caption{$c_4^q$, $T=0.1$, $\xi=3+1/4$}
        \label{fig:low_T_repulsive_alpha_non0_c4q_level2}
    \end{subfigure}
    \caption{Charge cumulants as a function of $\alpha$, probing Eqs.(\ref{eq:c2q_low_T_nonzero_alpha},\ref{eq:c4q_low_T_nonzero_alpha},\ref{eq:c2q_low_T_big_alpha}).}
    \label{fig:smallT_c2q}
\end{figure}

Another situation for which analytic results can be obtained is when the angle is finite, $0<\alpha<\pi/4$, and the temperature is low enough such that the dressed velocities of the magnons are smaller than $x/t=\tan\alpha$ (this always happens for low enough temperatures)\footnote{We are thankful to Alvise Bastianello for suggesting this.}. 
The details of the calculation for integer $\xi$ couplings can be found in Appendix \ref{app:finiteray}. The second cumulant of the topological charge is
\begin{equation}
    c_2^q = \sin\alpha \int \text{d}\theta \rho_S(\theta)\,,
    \label{eq:c2q_low_T_nonzero_alpha}
\end{equation}
which is compared to numerical results in Fig. \ref{fig:low_T_repulsive_alpha_non0_c2q}.

The fourth cumulant has a richer structure:
\begin{equation}
    c_4^q = \cos\alpha \bigg[ \int_{-\infty}^{\theta^*} \text{d}\theta \rho_S(\theta)(\tan\alpha-\tanh\theta) +
    5\int_{\theta^*}^{\infty} \text{d}\theta \rho_S(\theta)(\tanh\theta-\tan\alpha) \bigg]\,,
    \label{eq:c4q_low_T_nonzero_alpha}
\end{equation}
where $\tanh\theta^*=\tan\alpha$. This is compared to numerical results in Fig. \ref{fig:low_T_repulsive_alpha_non0_c4q}.

For $\alpha>\pi/4$, at low temperatures and $\xi$ a level-one repulsive coupling (see Appendix \ref{app:finiteray})
\begin{equation}
    F^q(\lambda) = \sin\alpha(\cosh\lambda -1) \int \text{d}\theta \rho_S(\theta)\,,
\end{equation}
from which it follows
\begin{equation}
    c_2^q = c_4^q = \sin\alpha \int \text{d}\theta \rho_S(\theta)\,,
    \label{eq:c2q_low_T_big_alpha}
\end{equation}
which is compared to numerics in Fig. \ref{fig:smallT_c2q}. We note that although the derivation applies strictly to level-1 reflective couplings, we also present results for level-2 couplings, for which the analytical results provide an accurate approximation as well. 
We furthermore remark that the approximation becomes unreliable at small ray angles that lie within the light cone set by the magnon velocities. These points will be discussed in more detail in Sec. \ref{sec:top_charge_fractal}.

\subsubsection{High temperature limit}
\label{sec:high_T_limit}

In the high temperature limit, we should recover the behaviour of the UV limit of the theory, which is the free massless boson conformal field theory. 

In this limit, the root densities become bimodal, concentrated around $\theta\approx \log(2T/m_S)$ corresponding to left and right movers. Extending the standard derivation employing Rogers dilogarithm identities to the case where there is a finite chemical potential $\mu$ for the topological charge, we obtain for the grand canonical free energy density over temperature
\begin{multline}
     \tilde f = -\sum_a\int \frac{\text{d}\theta}{2\pi}m_a\cosh\theta \log\left(1+e^{-\epsilon_a(\theta)}\right)\\
    =
     -\sum_a\int \frac{\text{d}\theta}{2\pi}\left[
     \frac{m_a}2e^\theta \log\left(1+e^{-\epsilon^\mathrm{R}_a(\theta)}\right)
     +\frac{m_a}2e^{-\theta} \log\left(1+e^{-\epsilon^\mathrm{L}_a(\theta)}\right)\right]=
    2\left(-\frac{\pi}{12R}-\frac{\bar \mu^2}{4\pi R}\Delta\right)\,,
    \label{eq:g}
\end{multline}
where $\epsilon^\mathrm{L,R}_a$ are the left and right halves of the pseudo-energies that are zero for $\theta>0$ and $\theta<0$, respectively, and in the result $R=1/T$, $\bar\mu=\mu/T$, and $\Delta=2\xi/(\xi+1) = \beta^2/(4\pi)$. The factor of 2 on the right-hand side shows that in equilibrium, the left and right contributions are equal. 

Due to the bimodal distributions in rapidity space, the flow equations \eqref{eq:flow_eq} decouple for the left and right movers, whose dressed velocities approach -1 and 1 (speed of light) independent of the state. As a consequence, the sign function becomes a constant along the flow. Assuming for simplicity $0\le\alpha\le\pi/2$ (i.e., $x,t>0$), it is $+1$ for left movers and also for right movers if $\tan\alpha>1$ (space-like separation), while it is $-1$ for right movers if $\tan\alpha<1$ (time-like separations). Using 
$h_a^{\text{dr}} = \eta_a\partial \epsilon_a/\partial \beta_j$,  the solution of the flow equation amounts to a shift in the chemical potentials $\beta_j$: $\epsilon_a(\theta,\lambda) = \epsilon_a(\theta)|_{\beta_j\to\beta_j
\pm\lambda}$ where the sign depends on $\eta_a$, the direction $\alpha$, and on the velocity ($v=\pm1$). 

In equilibrium ($\lambda=0$), the free energy \eqref{eq:g} is a sum of equal contributions of the left ($\epsilon^\mathrm{L}_a$) and right movers ($\epsilon^\mathrm{R}_a$). However, along the flow, their chemical potentials may get shifted in opposite directions due to the opposite signs of the velocities. For example, along the flow generated by the dressed energy, for $\tan\alpha<1$ we obtain
\begin{equation}
    \tilde f = -\frac{\pi}{12(R+\lambda)}-\frac{\bar \mu^2}{4\pi (R+\lambda)}\Delta -\frac{\pi}{12(R-\lambda)}-\frac{\bar \mu^2}{4\pi (R-\lambda)}\Delta\,.
\end{equation}
The expectation values of the conserved charges are generated as in Eq. \eqref{eq:fTderiv} at the shifted values of the chemical potentials (inverse temperature in this example).

As discussed above, the current densities are generated by the free energy flux \eqref{eq:flux} which is given by the same expression as $\tilde f$ but with $\cosh\theta$ replaced by $\sinh\theta$. The root densities are peaked at large rapidities, so we can replace the sinh function by a single exponential both for left and right movers, resulting in the integral expression in the second line of Eq. \eqref{eq:g} but with a $-$ sign between the two terms. Consequently, the current vanishes in equilibrium. However, along the flow, the two terms may become unequal, resulting in finite currents given by derivatives as in Eq. \eqref{eq:fTderiv}.

According to Eq. \eqref{eq:FCS}, for the FCS we need to integrate the charge and current densities along the flow with respect to $\lambda$. As lambda appears as a shift of the chemical potentials, the resulting expression will look very similar to the free energy. Taking into account all the above considerations, we obtain for the energy density
\begin{equation}
    \begin{aligned}
    F^e_\alpha(\lambda) &= \sin\alpha\,\frac{\pi}{6}
    \left(\frac1{R+\lambda}-\frac{1}{R}\right)&\quad x>t\,,\\
    F^e_\alpha(\lambda) &= \sin\alpha\,\frac{\pi}{12} \left(\frac1{R+\lambda}-\frac{1}{R-\lambda}\right)+\cos\alpha\,\frac{\pi}{12} \left(\frac1{R+\lambda}+\frac{1}{R-\lambda}-\frac{2}R\right)&\quad x<t\,.
\end{aligned}
\label{eq:Fe_highT}
\end{equation}
Numerical verification of these results is presented in Sec. \ref{sec:energy_and_momentum}.

For the topological charge, we obtain the simpler result
\begin{equation}
\begin{aligned}
    F^q_\alpha(\lambda) &= \sin\alpha\,\frac{\Delta}{2\pi R}
    \lambda^2&\quad x>t\,,\\
    F^q_\alpha(\lambda) &= \cos\alpha\,\frac{\Delta}{2\pi R}
    \lambda^2&\quad x<t\,.
\end{aligned}
\label{eq:Fq_highT}
\end{equation}
The quadratic $\lambda$-dependence implies that the topological charge fluctuations are Gaussian. Note that due to $\ell\sin\alpha=x$ and $\ell\cos\alpha=t$, we found that the charge distribution is independent of $t$ for space-like separations, and it is independent of $x$ for time-like separations. The first observation holds true for any conserved quantity, since for $x>t$ ($\alpha>\pi/4$), the chemical potentials in the left and right pseudo-energies are shifted in the same direction, resulting in vanishing currents all along the flow, which in turn implies $t$-independence for the FCS.

For the purely temporal case $\alpha=0$, the results in Eqs. \eqref{eq:Fe_highT} and \eqref{eq:Fq_highT} agree with the results derived for generic CFTs in Refs. \cite{2012JPhA...45J2001B,Bernard2014}\footnote{In the cited works, the authors studied non-equilibrium steady states after joining two semi-infinite regions with different temperatures and chemical potentials, but their results are also valid in equilibrium if we set the two states to be identical. Note that the U(1) (topological) charge there has a different normalisation.}. We conjecture that the equations \eqref{eq:Fe_highT} and \eqref{eq:Fq_highT} for $\alpha\neq0$ remain valid in generic CFTs if we multiply $F^e_\alpha(\lambda)$ by the central charge, which is 1 in our case. 

Finally, we observe that the result \eqref{eq:Fq_highT} for the topological charge can be understood from the two-point function of vertex operators in the free boson CFT. If we assume that, as long as we are interested in the large distance behaviour, we can rewrite the right-hand side of \eqref{eq:charfnasymp} as a two-point function, we obtain
\begin{equation}
    e^{\ell F^{q}_{\alpha}(\lambda)}\asymp \left\langle e^{i\nu\beta/(2\pi) \phi(x,t)} e^{-i\nu\beta/(2\pi) \phi(0,0)}\right\rangle\Big|_{\nu\to i\lambda}
    \sim\left(\frac{R^2}{\pi^2}\sinh\left[\frac\pi{R}(x-t)\right]\sinh\left[\frac\pi{R}(x+t)\right]\right)^{\lambda^2\beta^2/(16\pi^3)}\,.
\end{equation}
If $|x-t|,x+t\gg R=1/T$, the sinh functions can be approximated by exponentials, which yields Eq. \eqref{eq:Fq_highT}.

\begin{figure}[H]
    \centering
    \begin{subfigure}{0.24\textwidth}
        \includegraphics[height=2.5cm]{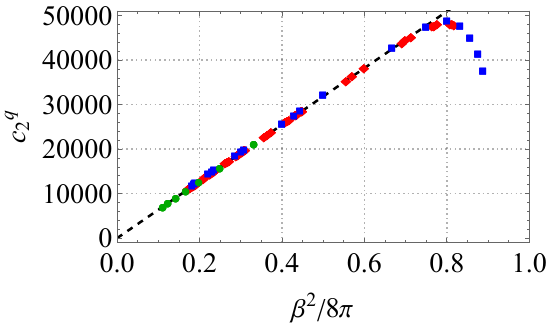}
        \caption{$T=10^5$, $\alpha=0$}
        \label{fig:high_T_reflectionless_alpha_0}
    \end{subfigure}
    \hfil
    \begin{subfigure}{0.24\textwidth}
        \includegraphics[height=2.5cm]{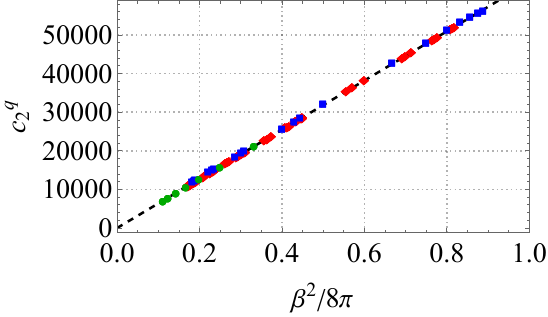}
        \caption{$T=10^5$, $\alpha=\pi/2$}
        \label{fig:high_T_reflectionless_alpha_pi_over_2}
    \end{subfigure}
    \begin{subfigure}{0.24\textwidth}
        \centering
        \includegraphics[height=2.5cm]{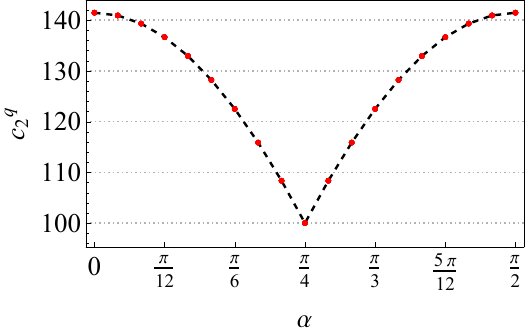}
        \caption{$T=1000.0$, $\xi=\frac{1}{3+1/2}$}
    \end{subfigure}
    \hfil
    \begin{subfigure}{0.24\textwidth}
        \centering
        \includegraphics[height=2.5cm]{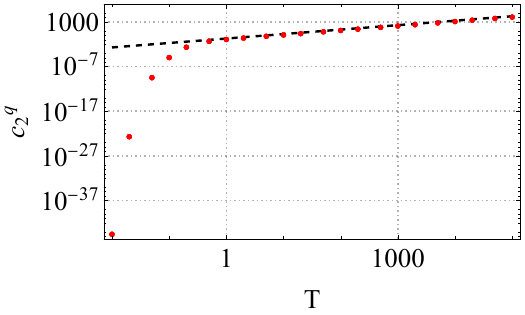}
        \caption{$\xi=\frac{1}{3+1/2}$, $\alpha=\pi/3$}
    \end{subfigure}
    
    \caption{High temperature limit of $c_2^q$ as a function of the coupling $\alpha$ and the temperature.}
    \label{fig:high_T}
\end{figure}

The only non-zero cumulant, $c_2^q$, is plotted together with the numerics in Fig. \ref{fig:high_T}, showing perfect agreement.

\subsection{Numerical results}
\subsubsection{Topological charge: fractal structure}
\label{sec:top_charge_fractal}

\begin{figure}[H]
    \centering
    
    \begin{subfigure}{0.24\textwidth}
        \centering
        \includegraphics[width=1\linewidth]{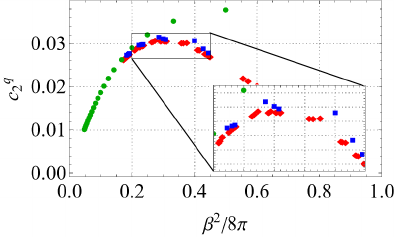}
        \caption{$\alpha=0$}
    \end{subfigure}
    \hfil
    \begin{subfigure}{0.24\textwidth}
        \centering
        \includegraphics[width=1\linewidth]{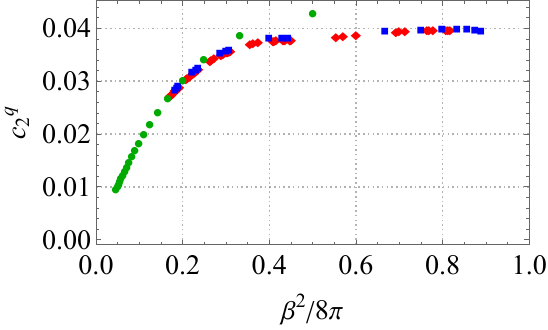}
        \caption{$\alpha=\frac{\pi}{6}$}
    \end{subfigure}
    \hfil
    \begin{subfigure}{0.24\textwidth}
        \centering
        \includegraphics[width=1\linewidth]{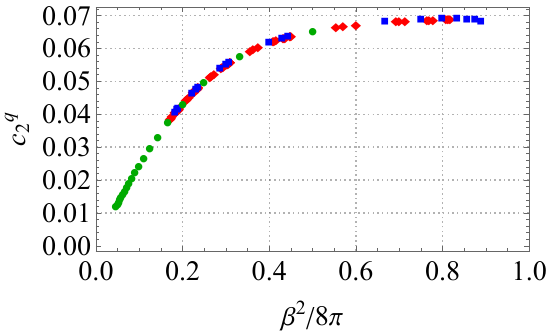}
        \caption{$\alpha=\frac{\pi}{3}$}
    \end{subfigure}
    \hfil
    \begin{subfigure}{0.24\textwidth}
        \centering
        \includegraphics[width=1\linewidth]{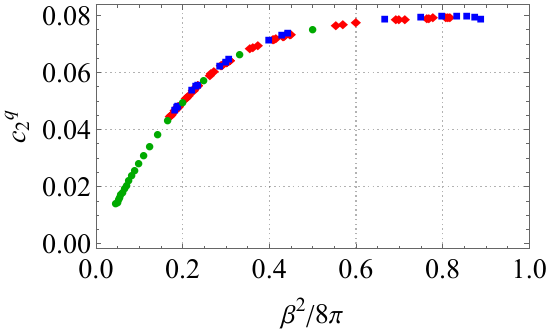}
        \caption{$\alpha=\frac{\pi}{2}$}
    \end{subfigure}    
    \caption{Variance $c_2^q$ for $T=0.5$ at different ray angles $\alpha$. The colours refer to the number of magnonic levels. For small angles, this quantity exhibits fractal behaviour, as shown in subfigures (a) and (b).}
     \label{fig:fractal}
\end{figure}

Due to the topological charge distribution being an even function at $\mu=0$, the odd cumulants of the topological charge are identically zero. The even cumulants show a fractal-like behaviour as a function of coupling up to some $\alpha$, which depends on the temperature, and become continuous above that $\alpha$. This is demonstrated in Fig. \ref{fig:fractal}. In contrast, the cumulants of the other quantities are continuous functions of the coupling, as demonstrated in the next subsection.

As discussed in Sec. \ref{sec:FCS}, the relevant random variable for the charge can be interpreted as the number of solitons crossing the world-line specified by the angle $\alpha$. Consequently, one generally expects any fractal features to vanish once $\alpha > \pi/4$, since no excitation can propagate outside the light cone. However, at low temperatures and for reflective values of the coupling, the effective light cone associated with charge transport is typically narrower \cite{2024PhRvB.109p1112M}. In such cases, the disappearance of the fractal structure is expected to occur already for $\alpha < \pi/4$.

\begin{figure}[H]
    \centering
    \includegraphics[height=3cm]{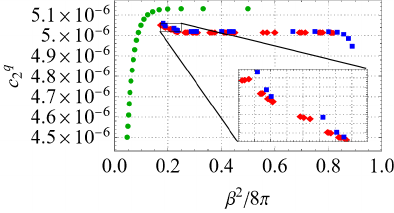}
    \hfil
    \includegraphics[height=3cm]{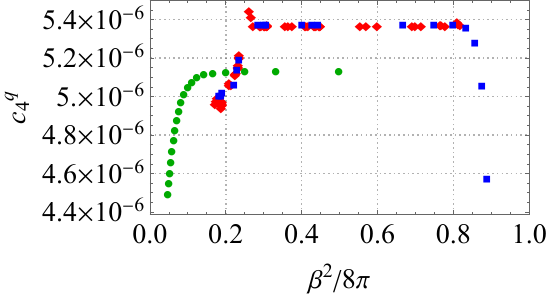}
    \hfil
    \includegraphics[height=3cm]{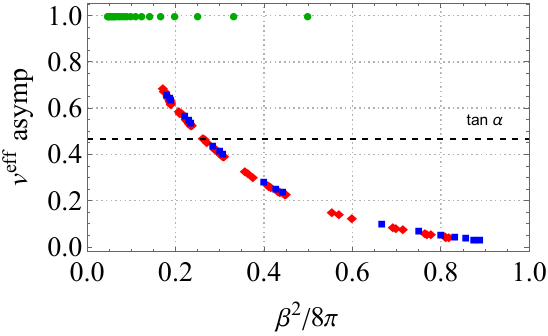}
    
    \includegraphics[height=3cm]{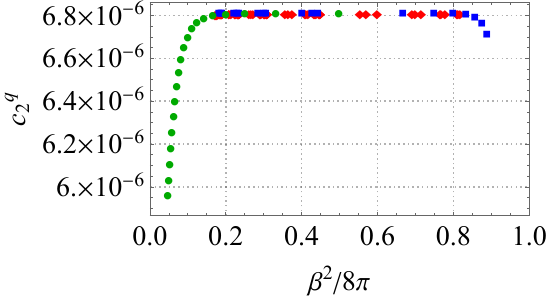}
    \hfil
    \includegraphics[height=3cm]{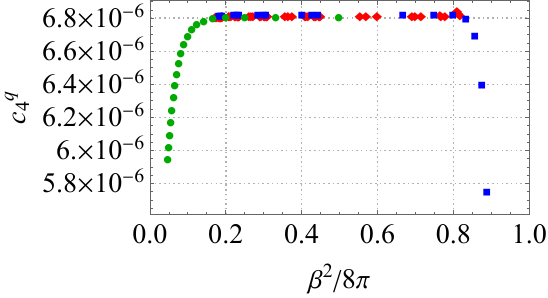}
    \hfil
    \includegraphics[height=3cm]{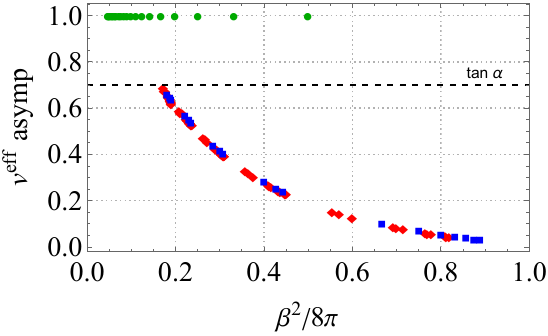}
    
    \caption{Cumulants $c_2^q$, $c_4^q$ and the asymptotic value of the last particle's (soliton in reflectionless and last magnon in reflective cases) effective velocity in terms of the coupling and two different rays, $\alpha=5\pi/36$ and $7\pi/36$, at $T=0.1$. For the smaller $\alpha$, the fractal structure is visible because magnons can propagate up to the world-line set by $\alpha$. For higher $\alpha$, the fractal structure disappears, because the magnons are too slow to affect the charge statistics.}
    \label{fig:fractal_disappearance}
\end{figure}

Figure \ref{fig:fractal_disappearance} displays $c_2^q$, $c_4^q$ and the maximum of the effective velocity of the last charge-carrying particle (the soliton at reflectionless points and the last magnon at reflective points) as functions of the coupling strength, for two different values of $\alpha$. At reflectionless couplings, charge degrees of freedom always propagate at a maximum velocity $1$, whereas for reflective couplings, the speed of charge propagation is reduced. When this reduced velocity exceeds $\tan\alpha$, magnons can propagate along the world line defined by $\alpha$, thereby influencing the charge statistics at these parameters and giving rise to the observed fractal behaviour. Conversely, when the magnonic velocities fall below $\tan \alpha$, these excitations cannot reach the world-line, and hence cannot affect the charge statistics. In this regime, the cumulants become continuous functions of the coupling strength.

\begin{figure}[H]
    \centering
    \includegraphics[height=4.9cm]{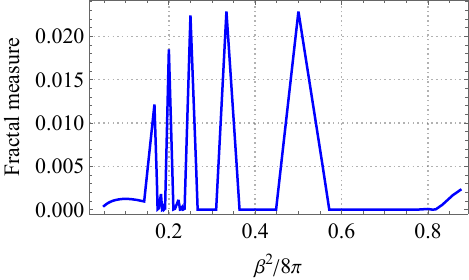}
    \hfil
    \includegraphics[height=5.1cm]{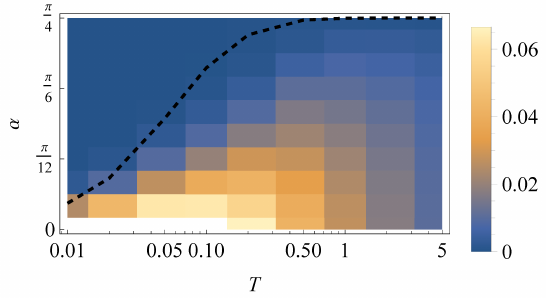}
    \caption{{\it Left:} An example of the fractal measure function  $fm(\xi_n;\alpha,T)$ in Eq.\ \eqref{eq:fractal_measure} at $T=0.1$ and $\alpha=5\pi/36$, the integral of which defines the proxy quantity shown in the right panel. 
    {\it Right:} Phase diagram constructed from this proxy quantity.  
    The dashed line marks the value of $\alpha$ at which fractal behaviour disappears for a given $T$ (cf. Fig. \ref{fig:fractal_disappearance}).}
    \label{fig:fractal_phase_diag}
\end{figure}

In Fig.~\ref{fig:fractal_phase_diag}, we map out the region of fractal behaviour in a sort of ``phase diagram'' in the $\alpha-T$ plane. To this end, we define a proxy fractal measure as follows. At fixed $T$ and $\alpha$ and for a fixed set of couplings $\{\xi_n\}$, we first define a function $fm(\xi_n;\alpha,T)$ which is equal at any $\xi=\xi_n$ to the relative difference between $c_2^q(\xi_n;\alpha,T)$ and the properly weighted average of $c_2^q(\xi_{n-1};\alpha,T)$ and $c_2^q(\xi_{n+1};\alpha,T)$,
\begin{equation}
\label{eq:fractal_measure}
fm(\xi_n;\alpha,T)=\left|\frac{c_2^q(\xi_n;\alpha,T)-\left(c_2^q(\xi_{n-1};\alpha,T)+\frac{\xi_n-\xi_{n-1}}{\xi_{n+1}-\xi_{n-1}}(c_2^q(\xi_{n+1};\alpha,T)-c_2^q(\xi_{n-1};\alpha,T))\right)}{c_2^q(\xi_n;\alpha,T)}\right|\,.
\end{equation}
For $\xi\neq\xi_n$, the function linearly interpolates between its values at neighbouring $\xi_n$ (see left panel of Fig.~\ref{fig:fractal_phase_diag}).
Integrating this function over $\xi$ yields the proxy quantity which is plotted in the phase diagram (right panel of Fig.~\ref{fig:fractal_phase_diag}) for a fixed $\alpha$ and $T$.

We note that more refined proxy quantities could be defined, but this would require substantially finer resolutions in $\xi$, at significantly greater computational cost. Furthermore, our proxy quantity is ``global'' in $\xi$ since the $\xi$-dependence is integrated out, whereas fractality is more naturally understood as a ``local'' property, depending explicitly on $\xi$. Despite these limitations, the phase diagram indicates that fractal behaviour is expected at low temperatures and small ray angles, and that the transition from fractal to non-fractal behaviour is not a sharp phase transition but rather a crossover. At high temperatures, our numerical results show that the fractal behaviour disappears at $\alpha=\pi/4$; however, this feature is not well captured by the proxy phase diagram.

These considerations explain the behaviour observed in Fig. \ref{fig:smallT_c2q} as well. At low temperatures and small ray angles, one generally expects fractal behaviour of the cumulants of the topological charge as a function of the coupling strength. The derivation in Sec. \ref{sec:level_one_repulsive}. relies on results valid only at one magnonic level, and such results are not expected to extend to couplings involving higher magnonic levels in the presence of fractal structures. However, as argued in this section, sufficiently large values of $\alpha$ suppress the fractal behaviour. Consequently, in Fig. \ref{fig:smallT_c2q}, good agreement with the analytical prediction is observed at larger ray angles even for couplings with two magnonic levels, while deviations appear at small ray angles, where the fractal structure persists.

\subsubsection{Energy and momentum}
\label{sec:energy_and_momentum}

In this section, we present our numerical results for the second cumulants of energy and momentum. These quantities are continuous functions of the coupling in all parameter regimes where we gathered data. This is in contrast to the cumulants of topological charge, which exhibit fractal behaviour in certain regimes. This also demonstrates the ability of our numerical methods to capture both fractal and non-fractal behaviours.

\begin{figure}[H]
    \centering
    
    \begin{subfigure}{0.24\textwidth}
        \centering
        \includegraphics[width=1\linewidth]{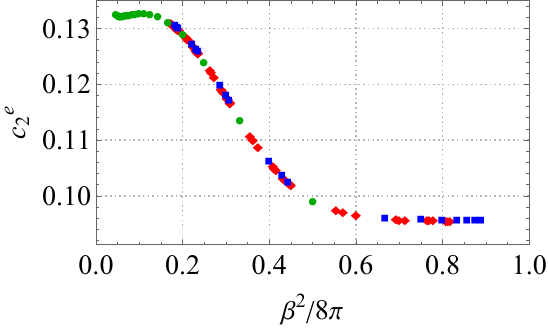}
    \end{subfigure}
    \hfil
    \begin{subfigure}{0.24\textwidth}
        \centering
        \includegraphics[width=1\linewidth]{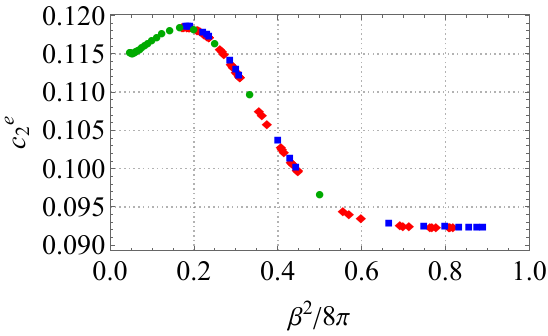}
    \end{subfigure}
    \hfil
    \begin{subfigure}{0.24\textwidth}
        \centering
        \includegraphics[width=1\linewidth]{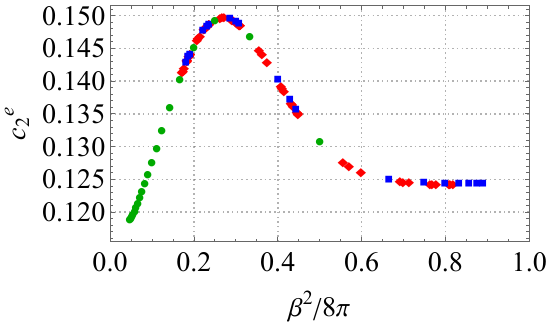}
    \end{subfigure}
    \hfil
    \begin{subfigure}{0.24\textwidth}
        \centering
        \includegraphics[width=1\linewidth]{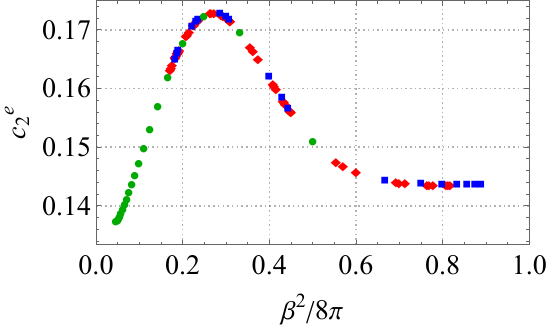}
    \end{subfigure}

    \begin{subfigure}{0.24\textwidth}
        \centering
        \includegraphics[width=1\linewidth]{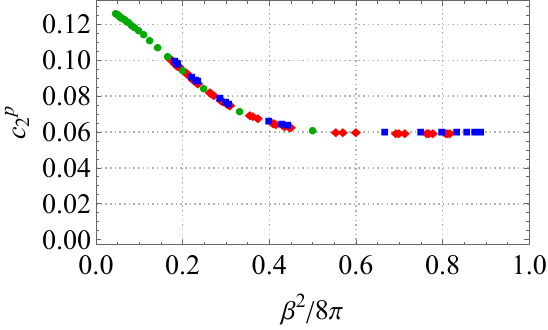}
        \caption{$\alpha=0$}
    \end{subfigure}
    \hfil
    \begin{subfigure}{0.24\textwidth}
        \centering
        \includegraphics[width=1\linewidth]{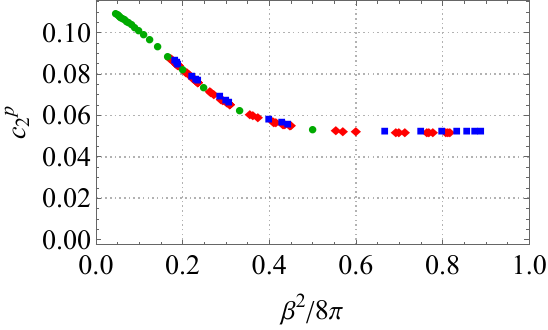}
        \caption{$\alpha=\frac{\pi}{6}$}
    \end{subfigure}
    \hfil
    \begin{subfigure}{0.24\textwidth}
        \centering
        \includegraphics[width=1\linewidth]{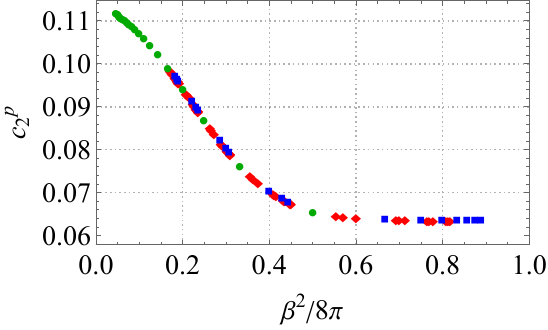}
        \caption{$\alpha=\frac{\pi}{3}$}
    \end{subfigure}
    \hfil
    \begin{subfigure}{0.24\textwidth}
        \centering
        \includegraphics[width=1\linewidth]{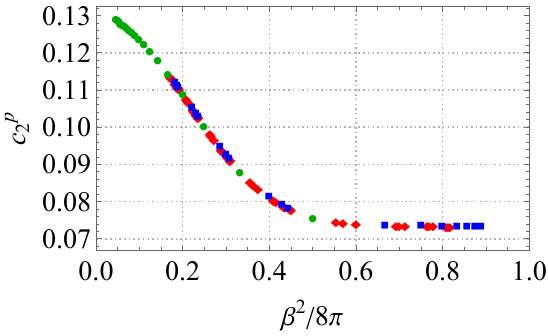}
        \caption{$\alpha=\frac{\pi}{2}$}
    \end{subfigure}
    
    \label{fig:placeholder}
    \caption{Variances $c_2^e$ and $c_2^p$ for $T=0.5$ at different ray angles $\alpha$. These quantities are not fractal at any of the parameters we studied.}
\end{figure}
\begin{figure}[H]
    \centering
    \begin{subfigure}{0.24\textwidth}
        \centering
        \includegraphics[height=2.35cm]{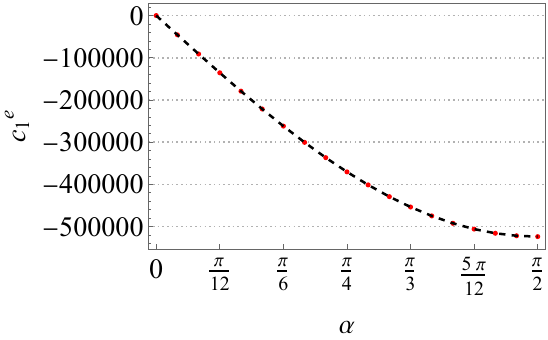}
    \end{subfigure}
    \hfil
    \begin{subfigure}{0.24\textwidth}
        \centering
        \includegraphics[height=2.35cm]{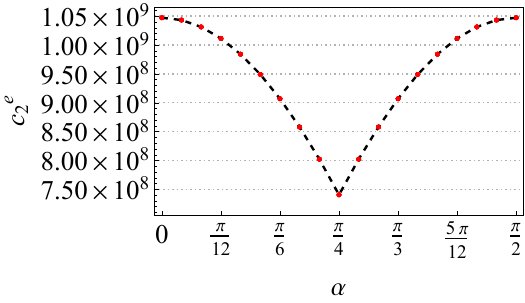}
    \end{subfigure}
    \hfil
    \begin{subfigure}{0.24\textwidth}
        \centering
        \includegraphics[height=2.35cm]{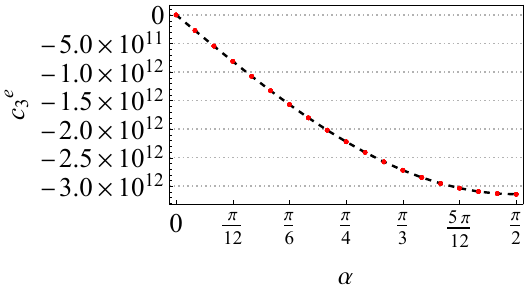}
    \end{subfigure}
    \hfil
    \begin{subfigure}{0.24\textwidth}
        \centering
        \includegraphics[height=2.35cm]{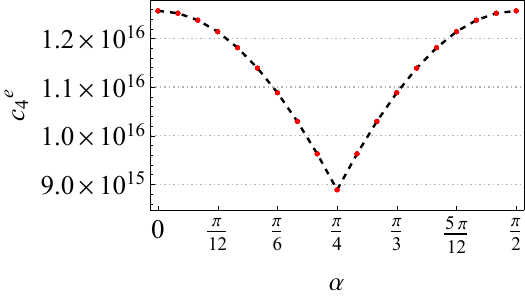}
    \end{subfigure}
    
    \begin{subfigure}{0.24\textwidth}
        \centering
        \includegraphics[height=2.35cm]{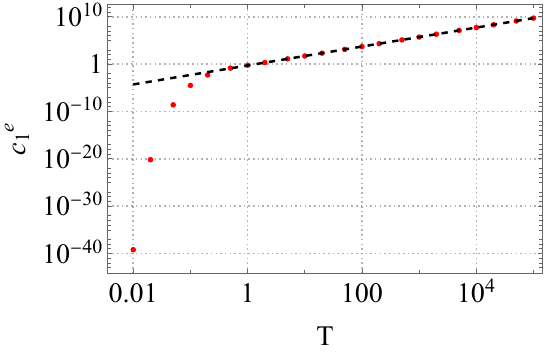}
        \caption{$c_1^e$}
    \end{subfigure}
    \hfil
    \begin{subfigure}{0.24\textwidth}
        \centering
        \includegraphics[height=2.35cm]{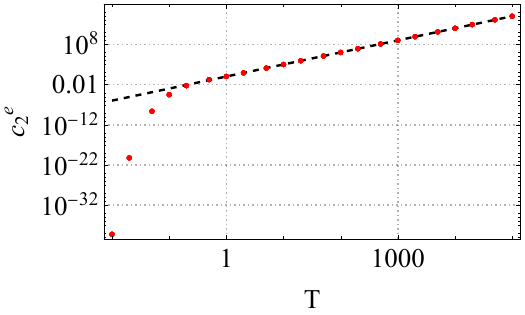}
        \caption{$c_2^e$}
    \end{subfigure}
    \hfil
    \begin{subfigure}{0.24\textwidth}
        \centering
        \includegraphics[height=2.35cm]{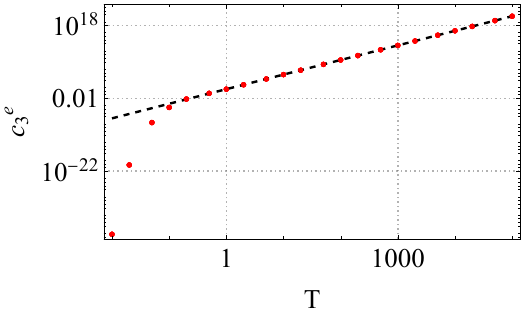}
        \caption{$c_3^e$}
    \end{subfigure}
    \hfil
    \begin{subfigure}{0.24\textwidth}
        \centering
        \includegraphics[height=2.35cm]{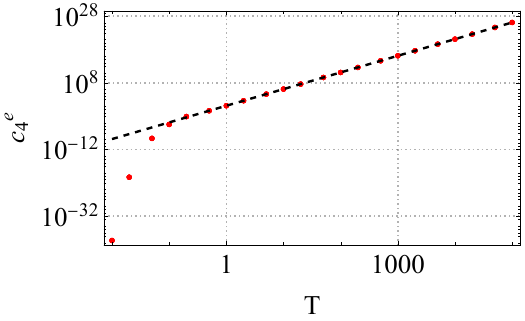}
        \caption{$c_4^e$}
    \end{subfigure}
    \caption{High temperature limit of the energy cumulants as a function of $\alpha$ (first row, $T=10^3$) and $T$ (second row, $\alpha=\pi/2$) The coupling is $\xi=\frac{1}{3+1/2}$ in each figure.}
    \label{fig:high_T_en}
\end{figure}
\begin{figure}[H]
    \centering
    \begin{subfigure}{0.32\textwidth}
        \centering
        \includegraphics[width=1\linewidth]{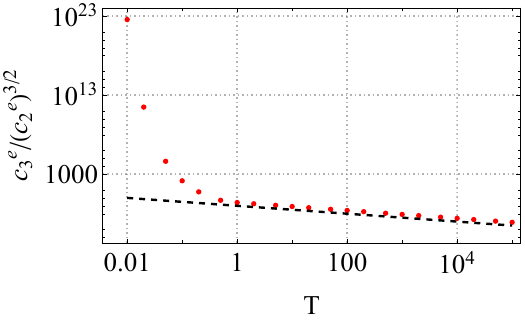}
        \caption{$\xi=1/3$, $\alpha=\pi/36$}
    \end{subfigure}
    \hfil
    \begin{subfigure}{0.32\textwidth}
        \centering
        \includegraphics[width=1\linewidth]{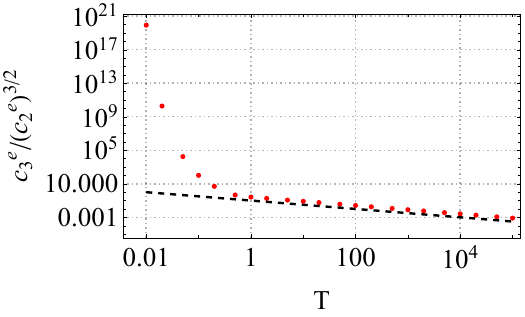}
        \caption{$\xi=\frac{1}{3+1/2}$, $\alpha=\pi/36$}
    \end{subfigure}
    \hfil
    \begin{subfigure}{0.32\textwidth}
        \centering
        \includegraphics[width=1\linewidth]{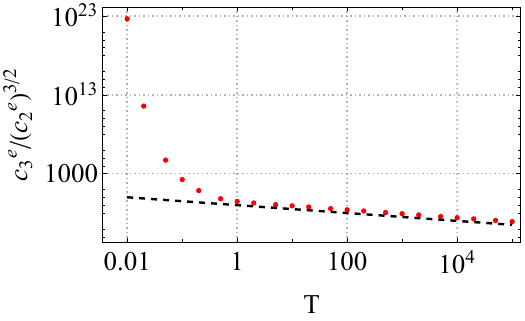}
        \caption{$\xi=3$, $\alpha=\pi/36$}
    \end{subfigure}

\vspace{0.3cm}

    \begin{subfigure}{0.32\textwidth}
        \centering
        \includegraphics[width=1\linewidth]{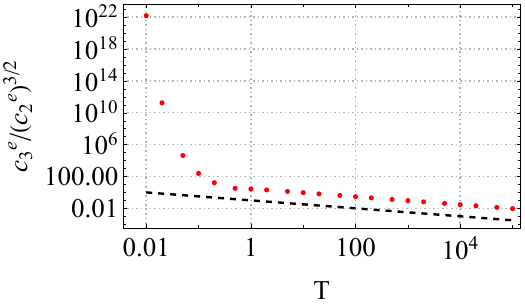}
        \caption{$\xi=1/3$, $\alpha=\pi/2$}
    \end{subfigure}
    \hfil
    \begin{subfigure}{0.32\textwidth}
        \centering
        \includegraphics[width=1\linewidth]{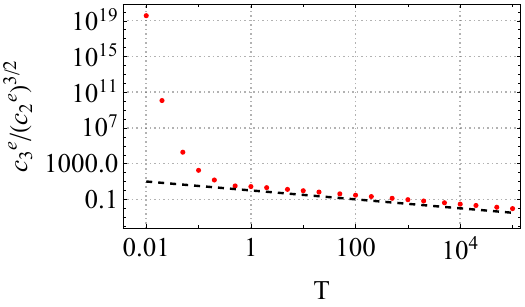}
        \caption{$\xi=\frac{1}{3+1/2}$, $\alpha=\pi/2$}
    \end{subfigure}
    \hfil
    \begin{subfigure}{0.32\textwidth}
        \centering
        \includegraphics[width=1\linewidth]{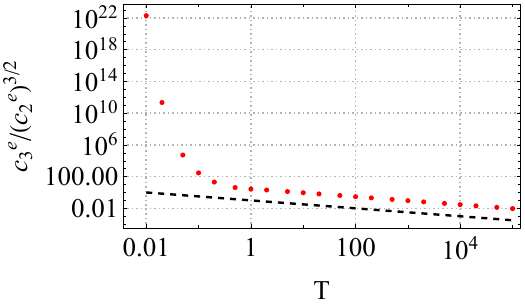}
        \caption{$\xi=3$, $\alpha=\pi/2$}
    \end{subfigure}
    \caption{Skewness of the energy-related distributions.}
    \label{fig:energy_skewness}
\end{figure}

From the results presented in Eq. \eqref{eq:Fe_highT}, one can extract the first four cumulants of the energy:
\begin{equation}
    c_1^e = -\frac{\pi T^2}{6} \sin\alpha\,,\qquad
    c_2^e=
    \begin{cases}
        \frac{\pi T^3}{3} \cos\alpha\,,\quad\alpha<\pi/4\\
        \frac{\pi T^3}{3} \sin\alpha\,,\quad\alpha>\pi/4
    \end{cases}
    \qquad
    c_3^e = -\pi T^4 \sin\alpha\,,\qquad
    c_4^e = 
    \begin{cases}
       4\pi T^5 \cos\alpha\,,\quad\alpha<\pi/4\\
       4\pi T^5\sin\alpha\,,\quad\alpha>\pi/4
    \end{cases}
\label{eq:en_high_T}
\end{equation}
The agreement with the numerics for a few parameters is shown in Fig. \ref{fig:high_T_en}.
From Eq.(\ref{eq:en_high_T}), one can see that in the high temperature limit, the skewness of the energy distribution is proportional to $T^{-1/2}$
\begin{equation}
    \frac{c_3^e}{(c_2^e)^{3/2}} \propto \frac{1}{\sqrt{T}}\,,
    \label{eq:skewness}
\end{equation}
which is verified numerically in Fig. \ref{fig:energy_skewness}.

\section{Conclusions}
\label{sec:conclusion}
In this work, we have conducted an extensive numerical analysis of the full counting statistics (FCS) of various conserved quantities and their currents in the sine--Gordon field theory. This study was initiated in Ref. \cite{DelVecchio2023}, which focused on the weak-coupling limit of the theory, which can be captured by the classical sine--Gordon model. We continued and considerably extended this work by exploring a wide range of temperatures, coupling strengths, and spatial and temporal directions.
Our investigation focused on the cumulants of the probability distributions associated with energy, momentum, and topological charge, as well as their corresponding currents. As the one-particle eigenvalues of other (higher) conserved charges are known \cite{Hegedus2026}, they could be studied in a similar way.

A central outcome of our study is the identification of a fractal dependence of the charge cumulants on the coupling strength, in contrast to the continuous behaviour observed for energy and momentum. This finding is consistent with, and extends, previous observations of fractal structures in transport quantities within the sine--Gordon model \cite{2025PhRvB.111k5121M}. The fact that only the topological charge displays fractal behaviour is due to the distinguished role of magnons in the dynamics of the topological charge, together with their complicated dependence on the coupling strength. We further provided evidence that this fractal behaviour disappears beyond the space-time rays past which charge-carrying excitations (i.e., magnons) cannot propagate at the given temperature.

From an analytical perspective, we have derived explicit expressions for the cumulants -- and, in specific regimes, the full distributions -- in several limiting cases. We validated these predictions by explicitly numerically computing the cumulants. 
In the high-temperature limit, we obtained analytic forms for the FCS of both energy and charge, recovering (and generalising) the expected conformal field theory results. 
We also demonstrated that the topological charge distribution derived from the BFT framework reproduces earlier semiclassical results \cite{2005PhRvL..95r7201D,PhysRevB.106.205151} in the low-temperature limit for reflectionless couplings. For generic couplings where the soliton-antisoliton scattering is non-diagonal and has a reflective component, our results differ from those of semiclassical approaches, not only in capturing the fractal behaviour, but also at low temperatures, where we derived explicit results for the cumulants of the topological charge. The semiclassical picture predicts \cite{PhysRevB.106.205151} anomalous, instead of ballistic, charge transport \cite{Gopalakrishnan2024,Krajnik2024} where the cumulants do not scale uniformly. This is a consequence of the predominantly reflective scatterings at low temperature \cite{2005PhRvL..95r7201D}. However, the scatterings are treated essentially classically, so the theory's integrability is not fully accounted for. It is an interesting question which approach describes a real system in which integrability is weakly broken, and on what time scales.

Although we have ensured that our numerical calculations are self-consistent and agree with the analytic limits, further confirmation via independent computational approaches would be highly desirable. In particular, methods capable of accessing either the cumulants directly or the relevant vertex-operator two-point functions would provide valuable external verification.

It would be even more compelling to test these results experimentally. 
A concrete protocol for measuring the distribution of the phase field --- closely related to the topological charge distribution --- as well as correlation functions in the sine--Gordon model has been proposed in \cite{DelVecchio2023}. 
This protocol, based on two tunnel-coupled quasi-condensates and matter-wave interferometry, is compatible with atom-chip implementations demonstrated by the Vienna group \cite{Schumm_2005,2007Natur.449..324H,10.21468/SciPostPhys.5.5.046}, which implements the sine--Gordon model in the weak-coupling regime ($\xi\lesssim0.1$). For such weak couplings, the fractal structure is likely to be suppressed below experimental resolution. Future experiments, such as those using fermionic atoms or coupled Hubbard chains \cite{Wybo2022,Wybo2023}, will realise the model at larger couplings, and our results can be used to benchmark these experiments.

It is important to note that the fractal structure appears in dynamical quantities, such as the cumulants of the integrated current, which require a two-time measurement scheme beyond destructive matter-wave interferometry. The experimental techniques based on generalised, partial outcoupling measurements are currently being developed \cite{2024PhRvL.133y0403P}, which will open the way to measuring dynamical correlations and distributions of time-integrated quantities.

\begin{acknowledgments}
MK would like to thank Alvise Bastianello, Benjamin Doyon, and Giuseppe Del Vecchio Del Vecchio for discussions and collaboration in the early stage of this work. This work was supported by the HUN-REN Hungarian Research Network through the Supported Research Groups Programme, HUN-REN-BME-BCE Quantum Technology Research Group (TKCS-2024/34), and also by the National Research, Development and Innovation Office (NKFIH) through the OTKA Grants K 138606 and ANN 142584. BN was partially supported by the Doctoral Excellence Fellowship Programme (DCEP), funded by the National Research, Development and Innovation Fund of the Ministry of Culture and Innovation, and the Budapest University of Technology and Economics, under a grant agreement with the National Research, Development and Innovation Office. GT was partially supported by the Quantum Information National Laboratory of Hungary (Grant No. 2022-2.1.1-NL-2022-00004).
\end{acknowledgments}

\bibliographystyle{utphys}
\bibliography{sg_fcs}

\providecommand{\href}[2]{#2}\begingroup\raggedright\begin{thebibliography}{10}

\bibitem{Touchette2009}
H.~Touchette, ``The large deviation approach to statistical mechanics,'' \href{http://dx.doi.org/https://doi.org/10.1016/j.physrep.2009.05.002}{{\em Physics Reports} {\bfseries 478} (2009) 1--69}, \href{http://arxiv.org/abs/0804.0327}{{\ttfamily arXiv:0804.0327 [cond-mat.stat-mech]}}.

\bibitem{Esposito2009}
M.~Esposito, U.~Harbola, and S.~Mukamel, ``Nonequilibrium fluctuations, fluctuation theorems, and counting statistics in quantum systems,'' \href{http://dx.doi.org/10.1103/RevModPhys.81.1665}{{\em Rev. Mod. Phys.} {\bfseries 81} (2009) 1665--1702}, \href{http://arxiv.org/abs/0811.3717}{{\ttfamily arXiv:0811.3717 [cond-mat.stat-mech]}}.

\bibitem{Bouchoule2006}
J.~Esteve, J.-B. Trebbia, T.~Schumm, A.~Aspect, C.~I. Westbrook, and I.~Bouchoule, ``Observations of density fluctuations in an elongated bose gas: Ideal gas and quasicondensate regimes,'' \href{http://dx.doi.org/10.1103/PhysRevLett.96.130403}{{\em Phys. Rev. Lett.} {\bfseries 96} (Apr, 2006) 130403}. \url{https://link.aps.org/doi/10.1103/PhysRevLett.96.130403}.

\bibitem{Bouchoule2010}
J.~Armijo, T.~Jacqmin, K.~V. Kheruntsyan, and I.~Bouchoule, ``Probing three-body correlations in a quantum gas using the measurement of the third moment of density fluctuations,'' \href{http://dx.doi.org/10.1103/PhysRevLett.105.230402}{{\em Phys. Rev. Lett.} {\bfseries 105} (Nov, 2010) 230402}. \url{https://link.aps.org/doi/10.1103/PhysRevLett.105.230402}.

\bibitem{Gritsev2007a}
V.~Gritsev, E.~Demler, M.~Lukin, and A.~Polkovnikov, ``Spectroscopy of collective excitations in interacting low-dimensional many-body systems using quench dynamics,'' \href{http://dx.doi.org/10.1103/PhysRevLett.99.200404}{{\em Phys. Rev. Lett.} {\bfseries 99} (2007) 200404}, \href{http://arxiv.org/abs/cond-mat/0702343}{{\ttfamily arXiv:cond-mat/0702343 [cond-mat.other]}}.

\bibitem{Schweigler2017}
T.~{Schweigler}, V.~{Kasper}, S.~{Erne}, I.~{Mazets}, B.~{Rauer}, F.~{Cataldini}, T.~{Langen}, T.~{Gasenzer}, J.~{Berges}, and J.~{Schmiedmayer}, ``{Experimental characterization of a quantum many-body system via higher-order correlations},'' \href{http://dx.doi.org/10.1038/nature22310}{{\em Nature} {\bfseries 545} (2017) 323--326}, \href{http://arxiv.org/abs/1505.03126}{{\ttfamily arXiv:1505.03126 [cond-mat.quant-gas]}}.

\bibitem{Wei2022}
D.~Wei, A.~Rubio-Abadal, B.~Ye, F.~Machado, J.~Kemp, K.~Srakaew, S.~Hollerith, J.~Rui, S.~Gopalakrishnan, N.~Y. Yao, I.~Bloch, and J.~Zeiher, ``Quantum gas microscopy of kardar-parisi-zhang superdiffusion,'' \href{http://dx.doi.org/10.1126/science.abk2397}{{\em Science} {\bfseries 376} (2022) 716--720}, \href{http://arxiv.org/abs/2107.00038}{{\ttfamily arXiv:2107.00038 [cond-mat.quant-gas]}}.

\bibitem{Fan2024}
Y.-Z. Fan and D.-B. Zhang, ``Full counting statistics of particle distribution on a digital quantum computer,'' \href{http://dx.doi.org/10.1103/PhysRevA.109.012412}{{\em Phys. Rev. A} {\bfseries 109} (2024) 012412}, \href{http://arxiv.org/abs/2308.01255}{{\ttfamily arXiv:2308.01255 [quant-ph]}}.

\bibitem{Rosenberg2024}
{E. Rosenberg et al.}, ``{Dynamics of magnetization at infinite temperature in a Heisenberg spin chain},'' \href{http://dx.doi.org/10.1126/science.adi7877}{{\em Science} {\bfseries 384} (2024) 48--53}, \href{http://arxiv.org/abs/2306.09333}{{\ttfamily arXiv:2306.09333 [quant-ph]}}.

\bibitem{Samajdar2024}
R.~Samajdar, E.~McCulloch, V.~Khemani, R.~Vasseur, and S.~Gopalakrishnan, ``Quantum turnstiles for robust measurement of full counting statistics,'' \href{http://dx.doi.org/10.1103/physrevlett.133.240403}{{\em Phys. Rev. Lett.} {\bfseries 133} (2024) 240403}, \href{http://arxiv.org/abs/2305.15464}{{\ttfamily arXiv:2305.15464 [quant-ph]}}.

\bibitem{Levitov1993}
L.~S. Levitov and G.~B. Lesovik, ``Charge distribution in quantum shot noise,'' {\em JETP Lett.} {\bfseries 58} (1993) 230.

\bibitem{Doyon2015}
B.~Doyon, A.~Lucas, K.~Schalm, and M.~J. Bhaseen, ``{Non-equilibrium steady states in the Klein–Gordon theory},'' \href{http://dx.doi.org/10.1088/1751-8113/48/9/095002}{{\em J. Phys. A Math. Theor.} {\bfseries 48} (2015) 095002}, \href{http://arxiv.org/abs/1409.6660}{{\ttfamily arXiv:1409.6660 [cond-mat.stat-mech]}}.

\bibitem{Yoshimura2018}
T.~Yoshimura, ``{Full counting statistics in the free Dirac theory},'' \href{http://dx.doi.org/10.1088/1751-8121/aae769}{{\em J. Phys. A Math. Theor.} {\bfseries 51} (2018) 475002}, \href{http://arxiv.org/abs/1802.01154}{{\ttfamily arXiv:1802.01154 [cond-mat.stat-mech]}}.

\bibitem{Bernard2016}
D.~Bernard and B.~Doyon, ``Conformal field theory out of equilibrium: a review,'' \href{http://dx.doi.org/10.1088/1742-5468/2016/06/064005}{{\em J. Stat. Mech. Theor. Exp.} {\bfseries 2016} (2016) 064005}, \href{http://arxiv.org/abs/1603.07765}{{\ttfamily arXiv:1603.07765 [cond-mat.stat-mech]}}.

\bibitem{Lamacraft2008}
A.~Lamacraft and P.~Fendley, ``{Order parameter statistics in the critical quantum Ising chain},'' \href{http://dx.doi.org/10.1103/physrevlett.100.165706}{{\em Phys. Rev. Lett.} {\bfseries 100} (2008) 165706}, \href{http://arxiv.org/abs/0802.1246}{{\ttfamily arXiv:0802.1246}}.

\bibitem{Stephan2017}
J.-M. Stéphan and F.~Pollmann, ``{Full counting statistics in the Haldane-Shastry chain},'' \href{http://dx.doi.org/10.1103/physrevb.95.035119}{{\em Phys. Rev. B} {\bfseries 95} (2017) 035119}, \href{http://arxiv.org/abs/1608.06856}{{\ttfamily arXiv:1608.06856 [cond-mat.str-el]}}.

\bibitem{Collura2017}
M.~Collura, F.~H.~L. Essler, and S.~Groha, ``{Full counting statistics in the spin-1/2 Heisenberg XXZ chain},'' \href{http://dx.doi.org/10.1088/1751-8121/aa87dd}{{\em J. Phys. A Math. Theor.} {\bfseries 50} (2017) 414002}, \href{http://arxiv.org/abs/1706.07939}{{\ttfamily arXiv:1706.07939 [cond-mat.stat-mech]}}.

\bibitem{Groha2018}
S.~Groha, F.~Essler, and P.~Calabrese, ``Full counting statistics in the transverse field ising chain,'' \href{http://dx.doi.org/10.21468/scipostphys.4.6.043}{{\em Scipost Phys.} {\bfseries 4} (2018) 043}, \href{http://arxiv.org/abs/1803.09755}{{\ttfamily arXiv:1803.09755 [cond-mat.stat-mech]}}.

\bibitem{Bastianello2018}
A.~{Bastianello}, L.~{Piroli}, and P.~{Calabrese}, ``{Exact Local Correlations and Full Counting Statistics for Arbitrary States of the One-Dimensional Interacting Bose Gas},'' \href{http://dx.doi.org/10.1103/PhysRevLett.120.190601}{{\em Phys. Rev. Lett.} {\bfseries 120} (2018) 190601}.

\bibitem{GangardtFCS2019}
M.~Arzamasovs and D.~M. Gangardt, ``Full counting statistics and large deviations in a thermal 1d bose gas,'' \href{http://dx.doi.org/10.1103/PhysRevLett.122.120401}{{\em Phys. Rev. Lett.} {\bfseries 122} (Mar, 2019) 120401}. \url{https://link.aps.org/doi/10.1103/PhysRevLett.122.120401}.

\bibitem{Perfetto2019}
G.~Perfetto, L.~Piroli, and A.~Gambassi, ``{Quench action and large deviations: Work statistics in the one-dimensional Bose gas},'' \href{http://dx.doi.org/10.1103/PhysRevE.100.032114}{{\em Phys. Rev. E} {\bfseries 100} (2019) 032114}, \href{http://arxiv.org/abs/1904.06259}{{\ttfamily arXiv:1904.06259 [cond-mat.stat-mech]}}.

\bibitem{Ares2021}
F.~Ares, M.~A. Rajabpour, and J.~Viti, ``{Exact full counting statistics for the staggered magnetization and the domain walls in the XY spin chain},'' \href{http://dx.doi.org/10.1103/physreve.103.042107}{{\em Phys. Rev. E} {\bfseries 103} (2021) 042107}, \href{http://arxiv.org/abs/2012.14012}{{\ttfamily arXiv:2012.14012 [cond-mat.stat-mech]}}.

\bibitem{Krajnik2024}
{\v{Z}}.~Krajnik, J.~Schmidt, V.~Pasquier, T.~Prosen, and E.~Ilievski, ``Universal anomalous fluctuations in charged single-file systems,'' \href{http://dx.doi.org/10.1103/physrevresearch.6.013260}{{\em Phys. Rev. Research} {\bfseries 6} (2024) 013260}, \href{http://arxiv.org/abs/2208.01463}{{\ttfamily arXiv:2208.01463 [cond-mat.stat-mech]}}.

\bibitem{Valli2025}
A.~Valli, C.~P. Moca, M.~A. Werner, M.~Kormos, {\v{Z}}.~Krajnik, T.~Prosen, and G.~Zar\'and, ``Efficient computation of cumulant evolution and full counting statistics: Application to infinite temperature quantum spin chains,'' \href{http://dx.doi.org/10.1103/f3c4-n21z}{{\em Phys. Rev. Lett.} {\bfseries 135} (2025) 100401}, \href{http://arxiv.org/abs/2409.14442}{{\ttfamily arXiv:2409.14442 [cond-mat.stat-mech]}}.

\bibitem{McCulloch2023}
E.~McCulloch, J.~De~Nardis, S.~Gopalakrishnan, and R.~Vasseur, ``Full counting statistics of charge in chaotic many-body quantum systems,'' \href{http://dx.doi.org/10.1103/PhysRevLett.131.210402}{{\em Phys. Rev. Lett.} {\bfseries 131} (2023) 210402}, \href{http://arxiv.org/abs/2302.01355}{{\ttfamily arXiv:2302.01355 [quant-ph]}}.

\bibitem{Gopalakrishnan2024}
S.~Gopalakrishnan, A.~Morningstar, R.~Vasseur, and V.~Khemani, ``Distinct universality classes of diffusive transport from full counting statistics,'' \href{http://dx.doi.org/10.1103/PhysRevB.109.024417}{{\em Phys. Rev. B} {\bfseries 109} (2024) 024417}, \href{http://arxiv.org/abs/2203.09526}{{\ttfamily arXiv:2203.09526 [cond-mat.stat-mech]}}.

\bibitem{Myers2020}
J.~Myers, J.~Bhaseen, R.~J. Harris, and B.~Doyon, ``{Transport fluctuations in integrable models out of equilibrium},'' \href{http://dx.doi.org/10.21468/SciPostPhys.8.1.007}{{\em SciPost Phys.} {\bfseries 8} (2020) 007}, \href{http://arxiv.org/abs/1812.02082}{{\ttfamily arXiv:1812.02082 [cond-mat.stat-mech]}}.

\bibitem{Doyon2019b}
B.~Doyon and J.~Myers, ``{Fluctuations in Ballistic Transport from Euler Hydrodynamics},'' \href{http://dx.doi.org/10.1007/s00023-019-00860-w}{{\em Ann. Henri Poincar{\'{e}}} {\bfseries 21} (2020) 255--302}, \href{http://arxiv.org/abs/1902.00320}{{\ttfamily arXiv:1902.00320}}.

\bibitem{Giamarchi:743140}
T.~Giamarchi, \href{http://dx.doi.org/10.1093/acprof:oso/9780198525004.001.0001}{{\em {Quantum physics in one dimension}}}.
\newblock International series of monographs on physics. Clarendon Press, Oxford, 2004.

\bibitem{Essler2005}
F.~H.~L. {Essler} and R.~M. {Konik}, \href{http://dx.doi.org/10.1142/9789812775344_0020}{``{Application of Massive Integrable Quantum Field Theories to Problems in Condensed Matter Physics},''} in {\em From Fields to Strings: Circumnavigating Theoretical Physics: Ian Kogan Memorial Collection (in 3 Vols)}, {M. Shifman et al.}, ed., pp.~684--830.
\newblock {World Scientific}, 2005.
\newblock \href{http://arxiv.org/abs/cond-mat/0412421}{{\ttfamily arXiv:cond-mat/0412421 [cond-mat.str-el]}}.

\bibitem{Zvyagin2004}
S.~A. Zvyagin, A.~K. Kolezhuk, J.~Krzystek, and R.~Feyerherm, ``Excitation hierarchy of the quantum sine-gordon spin chain in a strong magnetic field,'' \href{http://dx.doi.org/10.1103/PhysRevLett.93.027201}{{\em Phys. Rev. Lett.} {\bfseries 93} (2004) 027201}, \href{http://arxiv.org/abs/cond-mat/0403364}{{\ttfamily arXiv:cond-mat/0403364 [cond-mat.str-el]}}.

\bibitem{Controzzi2001}
D.~Controzzi, F.~H.~L. Essler, and A.~M. Tsvelik, \href{http://dx.doi.org/10.1007/978-94-010-0838-9_2}{``{Dynamical Properties of One Dimensional Mott Insulators},''} in {\em New Theoretical Approaches to Strongly Correlated Systems}, A.~M. Tsvelik, ed., p.~25–46.
\newblock Springer Netherlands, Dordrecht, 2001.
\newblock \href{http://arxiv.org/abs/cond-mat/0011439}{{\ttfamily arXiv:cond-mat/0011439 [cond-mat.str-el]}}.

\bibitem{Gritsev2007}
V.~{Gritsev}, A.~{Polkovnikov}, and E.~{Demler}, ``{Linear response theory for a pair of coupled one-dimensional condensates of interacting atoms},'' \href{http://dx.doi.org/10.1103/PhysRevB.75.174511}{{\em Phys. Rev. B} {\bfseries 75} (2007) 174511}, \href{http://arxiv.org/abs/cond-mat/0701421}{{\ttfamily arXiv:cond-mat/0701421 [cond-mat.other]}}.

\bibitem{Cirac2010}
J.~I. {Cirac}, P.~{Maraner}, and J.~K. {Pachos}, ``{Cold Atom Simulation of Interacting Relativistic Quantum Field Theories},'' \href{http://dx.doi.org/10.1103/PhysRevLett.105.190403}{{\em Phys. Rev. Lett.} {\bfseries 105} (2010) 190403}, \href{http://arxiv.org/abs/1006.2975}{{\ttfamily arXiv:1006.2975 [cond-mat.str-el]}}.

\bibitem{Haller2010}
E.~{Haller}, R.~{Hart}, M.~J. {Mark}, J.~G. {Danzl}, L.~{Reichs{\"o}llner}, M.~{Gustavsson}, M.~{Dalmonte}, G.~{Pupillo}, and H.-C. {N{\"a}gerl}, ``{Pinning quantum phase transition for a Luttinger liquid of strongly interacting bosons},'' \href{http://dx.doi.org/10.1038/nature09259}{{\em Nature} {\bfseries 466} (2010) 597--600}, \href{http://arxiv.org/abs/1004.3168}{{\ttfamily arXiv:1004.3168 [cond-mat.quant-gas]}}.

\bibitem{Wybo2023}
E.~Wybo, A.~Bastianello, M.~Aidelsburger, I.~Bloch, and M.~Knap, ``Preparing and analyzing solitons in the sine-gordon model with quantum gas microscopes,'' \href{http://dx.doi.org/10.1103/PRXQuantum.4.030308}{{\em PRX Quantum} {\bfseries 4} (2023) 030308}, \href{http://arxiv.org/abs/2303.16221}{{\ttfamily arXiv:2303.16221 [cond-mat.quant-gas]}}.

\bibitem{Wybo2022}
E.~Wybo, M.~Knap, and A.~Bastianello, ``{Quantum sine-Gordon dynamics in coupled spin chains},'' \href{http://dx.doi.org/10.1103/PhysRevB.106.075102}{{\em Phys. Rev. B} {\bfseries 106} (2022) 075102}, \href{http://arxiv.org/abs/2203.09530}{{\ttfamily arXiv:2203.09530}}.

\bibitem{Roy2021}
A.~{Roy}, D.~{Schuricht}, J.~{Hauschild}, F.~{Pollmann}, and H.~{Saleur}, ``{The quantum sine-Gordon model with quantum circuits},'' \href{http://dx.doi.org/10.1016/j.nuclphysb.2021.115445}{{\em Nucl. Phys. B} {\bfseries 968} (2021) 115445}, \href{http://arxiv.org/abs/2007.06874}{{\ttfamily arXiv:2007.06874 [quant-ph]}}.

\bibitem{Zache2020}
T.~V. Zache, T.~Schweigler, S.~Erne, J.~Schmiedmayer, and J.~Berges, ``Extracting the field theory description of a quantum many-body system from experimental data,'' \href{http://dx.doi.org/10.1103/PhysRevX.10.011020}{{\em Phys. Rev. X} {\bfseries 10} (2020) 011020}, \href{http://arxiv.org/abs/1909.12815}{{\ttfamily arXiv:1909.12815 [cond-mat.quant-gas]}}.

\bibitem{2024PhRvB.109c5118B}
A.~{Bastianello}, ``{Sine-Gordon model from coupled condensates: A generalized hydrodynamics viewpoint},'' \href{http://dx.doi.org/10.1103/PhysRevB.109.035118}{{\em Phys. Rev. B} {\bfseries 109} (2024) 035118}, \href{http://arxiv.org/abs/2310.04493}{{\ttfamily arXiv:2310.04493 [cond-mat.stat-mech]}}.

\bibitem{2023PhRvB.108x1105N}
B.~C. {Nagy}, M.~{Kormos}, and G.~{Tak{\'a}cs}, ``{Thermodynamics and fractal Drude weights in the sine-Gordon model},'' \href{http://dx.doi.org/10.1103/PhysRevB.108.L241105}{{\em Phys. Rev. B} {\bfseries 108} (2023) L241105}, \href{http://arxiv.org/abs/2305.15474}{{\ttfamily arXiv:2305.15474 [cond-mat.str-el]}}.

\bibitem{2024ScPP...16..145N}
B.~C. {Nagy}, G.~{Tak{\'a}cs}, and M.~{Kormos}, ``{Thermodynamic Bethe Ansatz and generalised hydrodynamics in the sine-Gordon model},'' \href{http://dx.doi.org/10.21468/SciPostPhys.16.6.145}{{\em SciPost Phys.} {\bfseries 16} (2024) 145}, \href{http://arxiv.org/abs/2312.03909}{{\ttfamily arXiv:2312.03909 [cond-mat.str-el]}}.

\bibitem{1987NuPhB.290..363D}
C.~{Destri} and H.~J. {De Vega}, ``{Light-cone lattice approach to fermionic theories in 2D The massive Thirring model},'' \href{http://dx.doi.org/10.1016/0550-3213(87)90193-3}{{\em Nucl. Phys. B} {\bfseries 290} (1987) 363--391}.

\bibitem{1995NuPhB.438..413D}
C.~{Destri} and H.~J. {de Vega}, ``{Unified approach to Thermodynamic Bethe Ansatz and finite size corrections for lattice models and field theories},'' \href{http://dx.doi.org/10.1016/0550-3213(94)00547-R}{{\em Nucl. Phys. B} {\bfseries 438} (1995) 413--454}, \href{http://arxiv.org/abs/hep-th/9407117}{{\ttfamily arXiv:hep-th/9407117 [hep-th]}}.

\bibitem{Hegedus2025}
{\'A}.~{Heged{\H{u}}s}, ``{Thermodynamics in the sine-Gordon model: the NLIE approach},'' \href{http://dx.doi.org/10.1016/j.nuclphysb.2025.117155}{{\em Nuclear Physics B} {\bfseries 1020} (2025) 117155}, \href{http://arxiv.org/abs/2507.19200}{{\ttfamily arXiv:2507.19200}}.

\bibitem{Hegedus2026}
{\'A}.~{Heged{\H{u}}s}, ``{NLIE formulations for the generalized Gibbs ensemble in the sine-Gordon model},'' \href{http://dx.doi.org/10.1016/j.nuclphysb.2026.117385}{{\em Nuclear Physics B} {\bfseries 1025} (2026) 117385}, \href{http://arxiv.org/abs/2510.25344}{{\ttfamily arXiv:2510.25344}}.

\bibitem{DelVecchio2023}
G.~{Del Vecchio Del Vecchio}, M.~{Kormos}, B.~{Doyon}, and A.~{Bastianello}, ``{Exact Large-Scale Fluctuations of the Phase Field in the Sine-Gordon Model},'' \href{http://dx.doi.org/10.1103/PhysRevLett.131.263401}{{\em Phys. Rev. Lett.} {\bfseries 131} (2023) 263401}, \href{http://arxiv.org/abs/2305.10495}{{\ttfamily arXiv:2305.10495 [cond-mat.stat-mech]}}.

\bibitem{koch2023exact}
R.~Koch and A.~Bastianello, ``{Exact Thermodynamics and Transport in the Classical Sine-Gordon Model},'' \href{http://dx.doi.org/10.21468/SciPostPhys.15.4.140}{{\em SciPost Phys.} {\bfseries 15} (2023) 140}, \href{http://arxiv.org/abs/2303.16932}{{\ttfamily arXiv:2303.16932 [cond-mat.stat-mech]}}.

\bibitem{2016PhRvX...6d1065C}
O.~A. {Castro-Alvaredo}, B.~{Doyon}, and T.~{Yoshimura}, ``{Emergent Hydrodynamics in Integrable Quantum Systems Out of Equilibrium},'' \href{http://dx.doi.org/10.1103/PhysRevX.6.041065}{{\em Phys. Rev. X} {\bfseries 6} (2016) 041065}, \href{http://arxiv.org/abs/1605.07331}{{\ttfamily arXiv:1605.07331 [cond-mat.stat-mech]}}.

\bibitem{2016PhRvL.117t7201B}
B.~{Bertini}, M.~{Collura}, J.~{De Nardis}, and M.~{Fagotti}, ``{Transport in Out-of-Equilibrium XXZ Chains: Exact Profiles of Charges and Currents},'' \href{http://dx.doi.org/10.1103/PhysRevLett.117.207201}{{\em Phys. Rev. Lett.} {\bfseries 117} (2016) 207201}, \href{http://arxiv.org/abs/1605.09790}{{\ttfamily arXiv:1605.09790 [cond-mat.stat-mech]}}.

\bibitem{2020PhRvX..10a1054B}
M.~{Borsi}, B.~{Pozsgay}, and L.~{Pristy{\'a}k}, ``{Current Operators in Bethe Ansatz and Generalized Hydrodynamics: An Exact Quantum-Classical Correspondence},'' \href{http://dx.doi.org/10.1103/PhysRevX.10.011054}{{\em Phys. Rev. X} {\bfseries 10} (2020) 011054}, \href{http://arxiv.org/abs/1908.07320}{{\ttfamily arXiv:1908.07320 [cond-mat.stat-mech]}}.

\bibitem{Krajnik2022}
{\v{Z}}.~Krajnik, J.~Schmidt, V.~Pasquier, E.~Ilievski, and T.~Prosen, ``Exact anomalous current fluctuations in a deterministic interacting model,'' \href{http://dx.doi.org/10.1103/PhysRevLett.128.160601}{{\em Phys. Rev. Lett.} {\bfseries 128} (2022) 160601}.

\bibitem{2022JSMTE2022e3102D}
G.~{Del Vecchio Del Vecchio} and B.~{Doyon}, ``{The hydrodynamic theory of dynamical correlation functions in the XX chain},'' \href{http://dx.doi.org/10.1088/1742-5468/ac6667}{{\em J. Stat. Mech. Theor. Exp.} {\bfseries 2022} (2022) 053102}, \href{http://arxiv.org/abs/2111.08420}{{\ttfamily arXiv:2111.08420 [math-ph]}}.

\bibitem{2005PhRvL..95r7201D}
K.~{Damle} and S.~{Sachdev}, ``{Universal Relaxational Dynamics of Gapped One-Dimensional Models in the Quantum Sine-Gordon Universality Class},'' \href{http://dx.doi.org/10.1103/PhysRevLett.95.187201}{{\em Phys. Rev. Lett.} {\bfseries 95} (2005) 187201}, \href{http://arxiv.org/abs/cond-mat/0507380}{{\ttfamily arXiv:cond-mat/0507380 [cond-mat.str-el]}}.

\bibitem{PhysRevB.106.205151}
M.~Kormos, D.~V\"or\"os, and G.~Zar\'and, ``{Finite-temperature dynamics in gapped one-dimensional models in the sine-Gordon family},'' \href{http://dx.doi.org/10.1103/PhysRevB.106.205151}{{\em Phys. Rev. B} {\bfseries 106} (2022) 205151}, \href{http://arxiv.org/abs/2208.08406}{{\ttfamily arXiv:2208.08406 [cond-mat.stat-mech]}}.

\bibitem{2012JPhA...45J2001B}
D.~{Bernard} and B.~{Doyon}, ``{Energy flow in non-equilibrium conformal field theory},'' \href{http://dx.doi.org/10.1088/1751-8113/45/36/362001}{{\em J. Phys. A Math. Gen.} {\bfseries 45} (2012) 362001}, \href{http://arxiv.org/abs/1202.0239}{{\ttfamily arXiv:1202.0239 [cond-mat.str-el]}}.

\bibitem{Bernard2014}
D.~Bernard and B.~Doyon, ``{Non-Equilibrium Steady States in Conformal Field Theory},'' \href{http://dx.doi.org/10.1007/s00023-014-0314-8}{{\em Ann. Henri Poincar{\'{e}}} (2014) 45}, \href{http://arxiv.org/abs/1302.3125}{{\ttfamily arXiv:1302.3125 [math-ph]}}.

\bibitem{2024PhRvB.109p1112M}
F.~{M{\o}ller}, B.~C. {Nagy}, M.~{Kormos}, and G.~{Tak{\'a}cs}, ``{Dynamical separation of charge and energy transport in one-dimensional Mott insulators},'' \href{http://dx.doi.org/10.1103/PhysRevB.109.L161112}{{\em Phys. Rev. B} {\bfseries 109} (2024) L161112}, \href{http://arxiv.org/abs/2311.16234}{{\ttfamily arXiv:2311.16234 [cond-mat.str-el]}}.

\bibitem{2025PhRvB.111k5121M}
F.~{M{\o}ller}, B.~C. {Nagy}, M.~{Kormos}, and G.~{Tak{\'a}cs}, ``{Anomalous charge transport in the sine-Gordon model},'' \href{http://dx.doi.org/10.1103/PhysRevB.111.115121}{{\em Phys. Rev. B} {\bfseries 111} (2025) 115121}, \href{http://arxiv.org/abs/2411.11473}{{\ttfamily arXiv:2411.11473 [cond-mat.stat-mech]}}.

\bibitem{Schumm_2005}
T.~{Schumm}, S.~{Hofferberth}, L.~M. {Andersson}, S.~{Wildermuth}, S.~{Groth}, I.~{Bar-Joseph}, J.~{Schmiedmayer}, and P.~{Kr{\"u}ger}, ``{Matter-wave interferometry in a double well on an atom chip},'' \href{http://dx.doi.org/10.1038/nphys125}{{\em Nature Physics} {\bfseries 1} (2005) 57--62}, \href{http://arxiv.org/abs/quant-ph/0507047}{{\ttfamily arXiv:quant-ph/0507047 [quant-ph]}}.

\bibitem{2007Natur.449..324H}
S.~{Hofferberth}, I.~{Lesanovsky}, B.~{Fischer}, T.~{Schumm}, and J.~{Schmiedmayer}, ``{Non-equilibrium coherence dynamics in one-dimensional Bose gases},'' \href{http://dx.doi.org/10.1038/nature06149}{{\em Nature} {\bfseries 449} (2007) 324--327}, \href{http://arxiv.org/abs/0706.2259}{{\ttfamily arXiv:0706.2259 [cond-mat.other]}}.

\bibitem{10.21468/SciPostPhys.5.5.046}
Y.~D. {van Nieuwkerk}, J.~{Schmiedmayer}, and F.~{Essler}, ``{Projective phase measurements in one-dimensional Bose gases},'' \href{http://dx.doi.org/10.21468/SciPostPhys.5.5.046}{{\em SciPost Phys.} {\bfseries 5} (2018) 046}, \href{http://arxiv.org/abs/1806.02626}{{\ttfamily arXiv:1806.02626 [cond-mat.quant-gas]}}.

\bibitem{2024PhRvL.133y0403P}
M.~{Pr{\"u}fer}, Y.~{Minoguchi}, T.~{Zhang}, Y.~{Kuriatnikov}, M.~V. {Ramana}, and J.~{Schmiedmayer}, ``{Quantum-Limited Generalized Measurement for Tunnel-Coupled Condensates},'' \href{http://dx.doi.org/10.1103/PhysRevLett.133.250403}{{\em \prl} {\bfseries 133} (2024) 250403}, \href{http://arxiv.org/abs/2408.07002}{{\ttfamily arXiv:2408.07002}}.

\bibitem{Bertini:2019lzy}
B.~{Bertini}, L.~{Piroli}, and M.~{Kormos}, ``{Transport in the sine-Gordon field theory: From generalized hydrodynamics to semiclassics},'' \href{http://dx.doi.org/10.1103/PhysRevB.100.035108}{{\em Phys. Rev. B} {\bfseries 100} (2019) 035108}, \href{http://arxiv.org/abs/1904.02696}{{\ttfamily arXiv:1904.02696 [cond-mat.stat-mech]}}.

\end{thebibliography}\endgroup

\clearpage
\appendix

\section{TBA expressions of cumulants}
\label{sec:app_cumulants_tba}

In general, the cumulant generating function of a random variable $X$ is defined as
\begin{equation}
    F(\lambda) = \log\langle e^{\lambda X}\rangle = \sum_n c_n \frac{\lambda^n}{n!}\,.
\end{equation}
The cumulants can be expressed in terms of the moments of $X$:
\begin{equation}
\begin{split}
    c_1 &= \langle X\rangle \\
    c_2 &= \langle X^2\rangle - \langle X\rangle^2 \\
    c_3 &= \langle X^3\rangle - 3 \langle X^2\rangle \langle X\rangle + 2 \langle X\rangle^3 \\
    c_4 &= \langle X^4\rangle - 4\langle X^3\rangle\langle X\rangle + 12 \langle X^2\rangle\langle X\rangle^2 - 3 \langle X^2\rangle^2 -6\langle X\rangle^4
\end{split}
\end{equation}
To present the expressions for the cumulants in terms of TBA quantities, we introduce the following notations
\begin{equation}
    H:=h^{(i)\text{dr}}\,,\qquad
    s = \text{sign}\left(\cos\alpha\ v^{\text{eff}}-\sin\alpha \right)\,,
\end{equation}
\begin{equation}
    \tilde{f}=-\left(\frac{1}{f}\frac{\partial f}{\partial \epsilon} + 2f\right)\,, \qquad
    \hat{f} = -\left(\frac{1}{f\tilde{f}}\frac{\partial \left(f\tilde{f}\right)}{\partial \epsilon} + 3f\right)\,.
\end{equation}
Compared to the original derivation \cite{Myers2020}, the sine--Gordon model contains the particle signs $\eta$, which need to be carefully followed. The flow equation with the above notation, and $\eta$ is
\begin{equation}
    \partial_{\lambda}\epsilon = -s\eta H\,.
\end{equation}
A useful relation when taking the derivatives is
\begin{equation}
    \partial_{\lambda}X^{\text{dr}} = \left(\partial_{\lambda}X\right)^{\text{dr}} + \left(s f H X^{\text{dr}}\right)^{\text{dr}} - \eta s f H X^{\text{dr}}\,.
\end{equation}
Starting from the TBA expression for the FCS
\begin{equation}
\begin{split}
    F_{\alpha}(\lambda) = 
    \int \text{d}\lambda\  \cos\alpha \langle j\rangle_{\lambda} -\sin\alpha \langle h\rangle_{\lambda} =
    \int \text{d}\lambda \int \frac{\text{d}\theta}{2\pi} \left(\cos\alpha\ e' -\sin\alpha\ p'\right)\vartheta H\,,
    \label{eq:FCS_TBA}
\end{split}
\end{equation}
and using the above notation and relations, one can derive the first four cumulants
\begin{subequations}
\begin{align}
\begin{split}
    c_1&=\partial_{\lambda}F_{\alpha}(\lambda)\big\rvert_{\lambda=0} =   \cos\alpha \langle j\rangle -\sin\alpha \langle h\rangle = \int_a \frac{\text{d}\theta}{2\pi} \left(\cos\alpha\ e' -\sin\alpha\ p'\right)\vartheta H\,,
    \label{eq:c1}
\end{split}\\
\begin{split}
    c_2 &= \partial_{\lambda}^2F_{\alpha}(\lambda)\big\rvert_{\lambda=0} = 
    \int_a \text{d}\theta\ \rho f \left|\cos\alpha\ v^{\text{eff}} -\sin\alpha\right| H^2\,,
    \label{eq:c2}
\end{split}\\
\begin{split}
    c_3 &= \partial_{\lambda}^3F_{\alpha}(\lambda)\big\rvert_{\lambda=0}
    = \int_a \text{d}\theta\ \rho f \left|\cos\alpha\ v^{\text{eff}} -\sin\alpha\right|H\left[ 3\left(sfH^2\right)^{\text{dr}}+\eta s \tilde{f}H^2 \right]\,,
    \label{eq:c3}
\end{split}\\
\begin{split}
    c_4 &= \partial_{\lambda}^4F_{\alpha}(\lambda)\big\rvert_{\lambda=0} = \int_a \text{d}\theta\ \rho f \left|\cos\alpha\ v^{\text{eff}} -\sin\alpha\right|\times
    \\ &\times \bigg\{
    \tilde{f}\hat{f}H^4 +
    3 \left[\left(sfH^2\right)^{\text{dr}}\right]^2 +
    4 H \left( \eta f \tilde{f} H^3 \right)^{\text{dr}} + 
    6 \eta s \tilde{f} H^2 \left(sfH^2\right)^{\text{dr}} + 
    12 H \left[sfH\left(sfH^2\right)^{\text{dr}}\right]^{\text{dr}}
    \bigg\}\,.
    \label{eq:c4}
\end{split}
\end{align}
\label{eq:app_cumulants}
\end{subequations}
When taking the derivatives, special care must be taken in treating the derivative of $s$. The original paper \cite{Myers2020} discusses the formally correct procedure, which effectively requires discarding all terms that explicitly involve derivatives of $s$.


\section{Low temperature limit in the repulsive regime}

In the repulsive regime, the only massive particle in the Bethe Ansatz solution is the soliton, which we label by the ``0'' subscript. In this section, we denote its mass by $M$ (i.e., $M=m_S$). The other excitations are the massless magnons. We mainly focus on the single-level case with $\xi$ magnon species, and on the topological charge, so we drop the `q' superscript for the sake of readability.

The TBA equations for the functions $\eta_a(\th)=e^{\epsilon_a(\th)}$ in thermal equilibrium in the partially decoupled form read as \cite{2023PhRvB.108x1105N,2024ScPP...16..145N}
\bes
\label{TBA2}
\begin{align}
\ln\eta_0&=\frac{M\cosh\th}T -s*\ln(1+\eta_1)\,,\label{eta0}\\
\ln\eta_1&=s*\ln\left[(1+\eta_0^{-1})(1+\eta_2)\right]+\delta_{\xi,3}s*\ln(1+\eta_\xi^{-1})\,,\label{eta1}\\
\ln\eta_n&=s*\ln\left[(1+\eta_{n-1})(1+\eta_{n+1})\right]\qquad\qquad 1<n<\xi-2\,,\\
\ln\eta_{\xi-2}&=s*\ln\left[(1+\eta_{\xi-3})(1+\eta_{\xi-1})(1+\eta_\xi^{-1})\right]\,,\\
\ln\eta_{\xi-1}&=\xi\frac\mu{T}+s*\ln(1+\eta_{\xi-2})\,,\label{eta_xim1_xi}\\
\ln\eta_\xi&=\xi\frac{\mu}{T}-s*\ln(1+\eta_{\xi-2})=2\xi\frac\mu{T}-\ln\eta_{\xi-1}
\,,\label{etap}
\end{align}
\esu
where $s(\th)=1/\cosh\th.$ For $\xi=2$ they are
\bes
\begin{align}
\ln\eta_0&=\frac{M\cosh\th}T -s*\ln[(1+\eta_1)(1+\eta_2^{-1})]\,,\\
\ln\eta_1&=\frac{2\mu}T+s*\ln(1+\eta_0^{-1})\,,\\
\ln\eta_2&=\frac{2\mu}T-s*\ln(1+\eta_0^{-1})=\frac{4\mu}T-\ln\eta_1\,.
\end{align}
\esu
In the grand canonical ensemble, $\varepsilon_{\xi-1}+\varepsilon_\xi=2\xi\mu/T.$

The dressed topological charges obey similar equations:
\bes
\label{hdr}
\begin{align}
h_0^\text{dr} &= -s\star(1-\vartheta_1)h_1^\text{dr}\,,\\    
h_1^\text{dr} &= s\star[(1-\vartheta_2)h_2^\text{dr}-\vartheta_0h_0^\text{dr}-\delta_{\xi,3}\vartheta_3h_3^\text{dr}]\,,\\
h_k^\text{dr} &= s\star[(1-\vartheta_{k-1})h_{k-1}^\text{dr}+
(1-\vartheta_{k+1})h_{k+1}^\text{dr}]\,,\\
h_{\xi-2}^\text{dr} &= s\star[(1-\vartheta_{\xi-3})h_{\xi-3}^\text{dr}+
(1-\vartheta_{\xi-1})h_{\xi-1}^\text{dr}-\vartheta_\xi h_\xi^\text{dr}]\,,\\
h_{\xi-1}^\text{dr} &= -\xi + s\star(1-\vartheta_{\xi-2})h_{\xi-2}^\text{dr}\,,\\
h_{\xi}^\text{dr} &= -\xi - s\star(1-\vartheta_{\xi-2})h_{\xi-2}^\text{dr}\,.
\end{align}
\esu
For $\xi=2,$
\bes
\label{hdrp2}
\begin{align}
h_0^\text{dr} &= -s\star(1-\vartheta_1)h_1^\text{dr}+s\star\vartheta_2 h_2^\text{dr}\,,\\    
h_1^\text{dr} &= -2 - s\star\vartheta_0 h_0^\text{dr}\,,\\
h_2^\text{dr} &= -2 + s\star\vartheta_0 h_0^\text{dr}\,.
\end{align}
\esu

In the limit $T\ll M,$ $\eta_0(\th)$ becomes exponentially large and can be neglected in the TBA equation for the magnons, so the magnonic TBA equations completely decouple from the kink equation. In a thermal state, the source terms are constant, so the solutions are also constant:
\bes
\begin{align}
&\eta_0(\theta) = \frac{e^{M\cosh\theta/T}}{2\cosh(\bar\mu)}\,,\\
&\eta_{k} = \!\!\left[\frac{\sinh[(k+1)\bar\mu ]}{\sinh(\bar\mu)}\right]^2\!\!\!-1\,,\,\,\quad1\leq k<\xi-1\,,\\
&\eta_{\xi-1}= e^{\xi \bar\mu}\frac{\sinh[( \xi -1)\bar\mu]}{\sinh(\bar\mu)}\,,\\
&\eta_\xi = e^{ \xi  \bar\mu}\frac{\sinh(\bar\mu)}{\sinh[( \xi -1)\bar\mu]}\,,
\end{align}
\label{eq:solution_etas}
\esu
where $\bar\mu=\mu/T.$ 

The root densities and dressed velocities can also be obtained in a closed form \cite{Bertini:2019lzy}. The kink root density is
\be
\label{rho0lowT}
\rho_0(\th)=\frac{M\cosh\th}{2\pi}2\cosh(\bar \mu) e^{-M\cosh\th/T}\,.
\ee
For the magnons, we quote the result for $\bar\mu=\mu/T=0,$
\bes
\label{rholimmu0}
\begin{align}
\rho_k(\theta) &= \frac{n}{4\pi} \frac1{k+1}\left(\frac{a_k(\theta)}{k}-\frac{a_{k+2}(\theta)}{k+2}\right)\,,\qquad k=1,\dots, \xi -2\,,\\ 
\rho_{ \xi -1}(\theta) &= \frac{n}{4\pi} \frac{a_{ \xi -1}(\theta)}{ \xi -1}\,,\\
\rho_{ \xi }(\theta) &= \frac{n}{4\pi} a_{ \xi -1}(\theta)\,,
\end{align}
\esu
where 
\be
n=\int\dd\th \rho_0(\th) = \sqrt{2MT/\pi}e^{-M/T}
\ee
is the kink density, and 
\be
a_k(\th) = \frac2 \xi  \frac{\sin\left(\frac{k\pi} \xi \right)}{\cosh\left(\frac{2\th} \xi \right)-\cos\left(\frac{k\pi} \xi \right)}\,.
\ee
%

Since all the root densities are $\sim e^{-M/T},$ we find that the dressed velocity of the kink is the bare velocity, $v_0(\th)=\tanh\th.$ This should hold true even in the $\lambda$-flow. In thermal equilibrium, the dressed velocities of the magnons can be obtained, for any $\mu$, in a way similar to the derivation of the $\rho_k$ with the result 
\be
\label{veff_lowT}
v_k(\th) \approx -\frac{T}M \frac{\rho'_k(\th)}{\rho_k(\th)}\,.
\ee

The generalised free energy (in thermal equilibrium) can be written as
\begin{multline}
\tilde f=\int\frac{\dd\th}{2\pi} M\cosh\theta\ln\left(1+2\cosh(\bar\mu)e^{-M\cosh\th/T}\right)
\approx
2\cosh(\bar\mu)\int\frac{\dd\th}{2\pi} M\cosh\theta e^{-M\cosh\th/T}=
\int\dd\th\rho_0(\th)=n\,,
\end{multline}
i.e., to leading order, it equals the kink density.

\subsection{Space-like case ($x>t,\,\alpha>\pi/4$)}

As discussed at the end of Sec. \ref{sec:FCS}, for space-like separations,
the FCS is given by the difference of generalised free energies. At low temperatures, this yields in the case of the topological charge, 
\be
F^q(\lambda) = \tilde f|_{\bar\mu\to \bar\mu-\lambda}-\tilde f|_{\bar\mu}=
2(\cosh \lambda-1)\int\frac{\dd\th}{2\pi} M\cosh\theta e^{-M\cosh\th/T} = 2\sinh^2(\lambda/2)\,n_{\bar\mu=0}\,,
\ee
so
\be
\big\langle e^{\lambda Q_x} \big\rangle\asymp
e^{-2\sinh^2(\lambda/2)nx}\,.
\ee
The result coincides with the one obtained for the reflectionless couplings in Sec. \ref{sec:refless_lowT}.

\subsection{Variance of the topological current}
\label{app:c2}

Let us turn now to the case $\alpha=0$ corresponding to the traditional full counting statistics. The second cumulant is
\be
c_2^q = \sum_{j=0}^ p  \int\dd\th \rho_j(\th)|v_j^\text{eff}(\th)|h_j^\text{dr}(\th)^2[1-\vartheta_j(\th)]\,.
\ee
For zero chemical potential, only the last two magnons have non-zero dressed topological charges equal to $\pm\xi$. Their filling fractions are constant (c.f. Eqs. \eqref{eq:solution_etas}). Because of Eq. \eqref{veff_lowT}, the integrand is essentially a total derivative, and we find
\be
\label{c2lowT}
c_2^q = 4 \xi \frac{T}M\frac{n}{4\pi}a_{ \xi -1}(0) = \frac{2nT}{\pi M}\tan\left(\frac\pi{2 \xi }\right) = \sqrt{\frac8M}\left(\frac{T}\pi\right)^{3/2}e^{-M/T}\tan\left(\frac\pi{2 \xi }\right)\,.
\ee

\subsection{Arbitrary ray ($\alpha\neq0$)}
\label{app:finiteray}

For any non-zero $\alpha$ and low enough temperatures, for the magnons $|v^\text{eff}_j(\theta,\lambda)|<\zeta=x/t$ will hold for all $\theta$ and $\lambda$. For simplicity, we assume $\zeta>0$. Then their flow equations become
\be
\partial_\lambda \varepsilon_k =  h_k^\text{dr}(\theta;\lambda)\,.
\ee
Using $h_k^\text{dr}=-\eta_k\partial_{\bar\mu}\varepsilon_k,$ 
this implies
\be
\varepsilon_k(\theta;\lambda) = \varepsilon_k(\theta;0)|_{h\to h-\eta_k\lambda}\,,
\ee
that is, the chemical potential gets shifted. For simplicity, we focus on the neutral case $\bar\mu=0$, but the derivation can be generalised to finite chemical potentials. Since at low $T$ the magnonic pseudo-energies are independent of the kink, they remain constant in $\theta$ along the flow and at each $\lam$ are given by the $\eps_k$ solutions in Eq. \eqref{eq:solution_etas} at $\bar\mu=-\lambda.$ 
Similarly, the dressed charges of the magnons are given by their derivatives with respect to $-\bar\mu$,  evaluated at $\bar\mu=-\lambda.$ 

The dressed charge of the kink can be obtained from the first equation of Eq. \eqref{hdr} or \eqref{hdrp2}, so it is also given by its equilibrium value at $\bar\mu=-\lambda,$
\be
h^\text{dr}_0(\theta;\lambda) =  
-\tanh(\lambda)\,.
\ee
Using again $v^\text{eff}_0(\th;\lam)=\tanh(\th)$, the flow equation for the kink is 
\be
\partial_\lambda\varepsilon_0(\theta;\lambda)=\mathrm{sign}(\tanh\theta-\zeta)\tanh(\lambda)\,.
\ee
The solution for $|\zeta|>1$ is
\be
\varepsilon_0(\theta;\lambda) = 
\varepsilon_0(\theta;0)-\ln\cosh(\lambda)\,,
\ee
while for $|\zeta|<1$ it is
\be
\varepsilon_0(\theta;\lambda) = 
\varepsilon_0(\theta;0)-\mathrm{sign}(\theta^*-\theta)\ln\cosh(\lambda)\,,
\ee
where $\tanh(\theta^*)=\zeta$. Note that both $h_0^\text{dr}$ and $\eps_0$ are independent of the coupling $\xi$.

We can now compute $F^q(\lam)$ using Eq. \eqref{eq:FCS}. 
For $|\zeta|>1$ we obtain
\begin{multline}
F^q(\lam) = \int_0^\lam \dd\lam' \int\frac{\dd\theta}{2\pi} 
(\cos\alpha\sinh\theta-\sin\alpha\cosh\theta)
2Me^{-M\cosh(\theta)/T}
\cosh(\lam')(-)\tanh\lam'\\
 =\cos\alpha \int\dd\theta\rho_0(\theta)(\zeta-\tanh\theta)(\cosh\lam-1)
=n\cos\alpha |\zeta|(\cosh\lam-1)
\,,\qquad |\zeta|>1\,,
\end{multline}
where we used Eq. \eqref{rho0lowT} with $h=0$ and that $\zeta=\tan\alpha,$ and then exploited that the odd $\tanh\theta$ term does not contribute and $n=\int\dd\theta\,\rho_0(\theta)$ is the total density of kinks. Then 
\be
-\ell F^q(\lam) = -\sqrt{x^2+t^2}\,F(\lam) = -2n\sinh^2(\lam/2)|x|\,, \qquad\qquad |\zeta|>1
\ee
which is the same as for the equal time ($\alpha=\pi/2$) case, as it should.

For $|\zeta|<1$ we obtain
\be
F^q(\lam) = \int_0^\lam \dd\lam' \int\frac{\dd\theta}{2\pi} 
(\cos\alpha\sinh\theta-\sin\alpha\cosh\theta)
2Me^{-M\cosh(\theta)/T}
(\cosh\lam')^{\mathrm{sgn}(\theta^*-\theta)}(-)\tanh\lam'\,.
\ee
Breaking up the integral into two pieces according to $\theta<\theta^*$ and $\theta>\theta^*$, the integral over $\lam$ can be performed giving
\be
F^q(\lam) = \cos\alpha \int_{-\infty}^{\theta^*}\dd\theta\rho_0(\theta)(\zeta-\tanh\theta)\left[(\cosh\lam)-1\right] +
\cos\alpha \int_{\theta^*}^\infty\dd\theta\rho_0(\theta)(\tanh\theta-\zeta)\left[(\cosh\lam)^{-1}-1\right]
\ee
which can also be written as
\be
F^q(\lam) =\cos\alpha \int\dd\theta\rho_0(\theta)\,|\zeta-\tanh\theta|\left[(\cosh\lam)^{\mathrm{sign}(\theta^*-\theta)}-1\right]\,,\qquad\qquad |\zeta|<1\,.
\ee
This implies that 
\be
-\ell F^q(\lam) = -\int\dd\theta\rho_0(\theta)\,|x-v_0(\th)t|\left[(\cosh\lam)^{\mathrm{sign}(\theta^*-\theta)}-1\right]\,.
\ee
We can compute the cumulants for $0<\zeta<1$:
\begin{multline}
F^q(\lam)|_{\lam=0} = 
\cos\alpha \int_{-\infty}^{\theta^*}\dd\theta\rho_0(\theta)(\zeta-\tanh\theta)
\left(\frac{\lambda^2}2+\frac{\lambda^4}{24}+\dots\right)+
\cos\alpha \int_{\theta^*}^\infty\dd\theta\rho_0(\theta)(\tanh\theta-\zeta)
\left(\frac{-\lambda^2}2+\frac{5\lambda^4}{24}+\dots\right)\\
=\lambda^2\cdot\zeta\cos\alpha \int\dd\theta\rho_0(\theta) +
\lambda^4\cdot\cos\alpha\left[ \int_{-\infty}^{\theta^*}\dd\theta\rho_0(\theta)(\zeta-\tanh\theta)+
5\int_{\theta^*}^\infty\dd\theta\rho_0(\theta)(\tanh\theta-\zeta)\right]+\dots,
\end{multline}
where we exploited that in the $O(\lambda^2)$ term, $\tanh(\theta)$ does not contribute being an odd function, and the two integrals can be merged. After multiplying by $\sqrt{x^2+t^2}$, we find that the second cumulant is simply $c_2=n|x|$ (the absolute value comes from considering $-1<\zeta<0$), which, interestingly, coincides with the space-like case. However, the higher cumulants are different.


\section{Further observations}
\label{sec:gen_obs}
In this section, we present our observations for the quantities that received less attention in the main text: $c_3^e$, $c_3^p$, $c_4^e$, $c_4^p$, $c_4^q$. Note that we omit $c_3^q$ here, because it is zero due to the symmetry of the topological charge distribution.

\begin{itemize}
    \item $c_3^e$, see Fig. \ref{fig:c3e}.
    \begin{itemize}
        \item $c_3^e$ is negative, and the graphs show its absolute value.
        \item Although the plots seem to show some structure for $\alpha=0$, the magnitudes show that $c_3^e$ is zero at $\alpha=0$, corresponding to the probability density function of the time-integrated energy current being an even function.
        \item As a function of $\alpha$, it's always a monotonically decreasing function, unlike the $c_2$ values.
        \item As a function of the coupling, it behaves similarly to the $c_2$ values: it goes to a constant for very small temperatures, then develops a local maximum, which is shifted toward higher couplings as the temperature increases.
    \end{itemize}
    \item $c_3^p$, see Fig. \ref{fig:c3p}.
    \begin{itemize}
        \item As a function of $\beta$ it shows the same characteristics as $c_2^p$, however, as a function of $\alpha$ it is monotonically decreasing and is a concave function.
        \item One can also see that for $\alpha=\pi/2$, it becomes zero, corresponding to the space-integrated momentum probability density function being an even function.
    \end{itemize}
    \item $c_4^e$ shows similar characteristics to $c_2^e$, see Fig. \ref{fig:c4e}.
    \item $c_4^p$ shows similar characteristics to $c_2^p$, see Fig. \ref{fig:c4p}.
    \item $c_4^q$, see Fig. \ref{fig:c4q}.
    \begin{itemize}
        \item We have shown in the main text that in the $T\to\infty$ limit, this quantity vanishes.
        Interestingly, for finite temperatures, we obtain finite values. The fractal structure can also be seen for the appropriate combinations of temperature and ray angle.
    \end{itemize}
\end{itemize}

\newpage
\cumulantplot{c3e}{Third cumulant of energy related distributions, $c_3^e$.}

\newpage
\cumulantplot{c3p}{Third cumulant of momentum related distributions, $c_3^p$.}

\newpage
\cumulantplot{c4e}{Fourth cumulant of energy related distributions, $c_4^e$.}

\newpage
\cumulantplot{c4p}{Fourth cumulant of momentum related distributions, $c_4^p$}

\newpage
\cumulantplot{c4q}{Fourth cumulant of topological charge related distributions, $c_4^q$}

\section{Comparison of numerical derivatives versus analytical expressions}
\label{app:comparison_subsec}

In this section, we compare the numerical values obtained by taking multiple numerical derivatives of $F$ with the analytical integral expressions. In the following three subsections, we present numerical results for the first four cumulants for the topological charge, the energy, and the momentum at temperature $T=1.0$, ray $\alpha=\pi/6$, and coupling $\xi=\frac{1}{3+1/2}$. In each case, the first table shows the FCS function and the cumulants calculated from their respective integral expressions. In the subsequent tables, we show the results of taking numerical derivatives from the values in the first table. For ease of comparison, the values corresponding to $\lambda=0$ in the first tables, and the first derivatives in the subsequent tables are presented in bold. The tables demonstrate that only the first numerical derivative can be computed with reliable precision; further derivatives fail to match the results from the integral expressions. The ``accuracy'' reported in the tables indicates the number of terms used to calculate the derivative using finite difference coefficients.

\subsection{Charge}

\begin{table}[H]
\centering
\caption*{Values of the charge full counting statistics $F$ and charge cumulants $c_i$ as functions of $\lambda$.}
\begin{tabular}{c|ccccccc}
\toprule
$\lambda$
& $-0.03$ & $-0.02$ & $-0.01$ & $0$ & $0.01$ & $0.02$ & $0.03$ \\
\midrule
$F$
& $4.42279\times10^{-5}$
& $1.96571\times10^{-5}$
& $4.91435\times10^{-6}$
& $0$
& $4.91435\times10^{-6}$
& $1.96571\times10^{-5}$
& $4.42279\times10^{-5}$ \\

$c_1$
& $-2.94846\times10^{-3}$
& $-1.96569\times10^{-3}$
& $-9.82871\times10^{-4}$
& $\mathbf{-5.64413\times10^{-18}}$
& $9.82871\times10^{-4}$
& $1.96569\times10^{-3}$
& $2.94846\times10^{-3}$ \\

$c_2$
& $0.0982870$
& $0.0982865$
& $0.0982876$
& $\mathbf{0.0982903}$
& $0.0982876$
& $0.0982865$
& $0.0982870$ \\

$c_3$
& $-4.83109\times10^{-4}$
& $-3.22075\times10^{-4}$
& $-1.61031\times10^{-4}$
& $\mathbf{9.80084\times10^{-18}}$
& $1.61031\times10^{-4}$
& $3.22075\times10^{-4}$
& $4.83109\times10^{-4}$ \\

$c_4$
& &
& &
$\mathbf{0.0161017}$
& &
& \\
\bottomrule
\end{tabular}
\end{table}

\begin{table}[H]
\centering

\begin{subtable}[t]{0.48\textwidth}
\vspace{0pt}
\centering
\caption*{$F$ derivatives}
\begin{tabular}{lccc}
\toprule
accuracy & 2 & 4 & 6 \\
\midrule
$c_1$ & $\mathbf{3.1552\times10^{-17}}$ & $\mathbf{2.56651\times10^{-17}}$ & $\mathbf{2.33781\times10^{-17}}$ \\
$c_2$ & 0.0982871 & 0.0982875 & 0.0982876 \\
$c_3$ & $3.51519\times10^{-13}$ & $5.2516\times10^{-13}$ & \\
$c_4$ & -0.0546484 & -0.0664486 & \\
\bottomrule
\end{tabular}
\end{subtable}
\hfill
\begin{subtable}[t]{0.48\textwidth}
\vspace{0pt}
\centering
\caption*{$c_1$ derivatives}
\begin{tabular}{lccc}
\toprule
accuracy & 2 & 4 & 6 \\
\midrule
$c_2$ & $\mathbf{0.0982871}$ & $\mathbf{0.098288}$ & $\mathbf{0.0982882}$ \\
$c_3$ & $1.48753\times10^{-12}$ & $1.86916\times10^{-12}$ & $1.88217\times10^{-12}$ \\
$c_4$ & -0.0546484 & -0.0723487 & \\
\bottomrule
\end{tabular}
\end{subtable}
\end{table}

\begin{table}[H]
\centering

\begin{subtable}[t]{0.48\textwidth}
\vspace{0pt}
\centering
\caption*{$c_2$ derivatives}
\begin{tabular}{lccc}
\toprule
accuracy & 2 & 4 & 6 \\
\midrule
$c_3$ & $\mathbf{0}$ & $\mathbf{6.93889\times10^{-16}}$ & $\mathbf{6.93889\times10^{-16}}$ \\
$c_4$ & -0.0546643 & -0.0664604 & 2675.61 \\
\bottomrule
\end{tabular}
\end{subtable}
\hfill
\begin{subtable}[t]{0.48\textwidth}
\vspace{0pt}
\centering
\caption*{$c_3$ derivatives}
\begin{tabular}{lccc}
\toprule
accuracy & 2 & 4 & 6 \\
\midrule
$c_4$ & $\mathbf{0.0161031}$ & $\mathbf{0.0161028}$ & $\mathbf{0.0161027}$ \\
\bottomrule
\end{tabular}
\end{subtable}

\end{table}

\subsection{Energy}

\begin{table}[H]
\centering
\caption*{Values of the energy full counting statistics $F$ and charge cumulants $c_i$ as functions of $\lambda$.}
\begin{tabular}{c|ccccccc}
\toprule
$\lambda$
& $-0.03$ & $-0.02$ & $-0.01$ & $0$ & $0.01$ & $0.02$ & $0.03$ \\
\midrule
$F$
& 0.00925624
& 0.00607612
& 0.0029913
& 0
& -0.00289949
& -0.00570876
& -0.00842926 \\

$c_1$
& -0.322822
& -0.313201
& -0.303763
& $\mathbf{-0.294497}$
& -0.285401
& -0.276454
& -0.267645 \\

$c_2$
& 0.971075
& 0.952207
& 0.934499
& $\mathbf{0.917915}$
& 0.902449
& 0.888042
& 0.874669 \\

$c_3$
& -1.94455
& -1.82663
& -1.71255
& $\mathbf{-1.60195}$
& -1.49454
& -1.38996
& -1.28788 \\

$c_4$
& &
& &
$\mathbf{10.8948}$
& &
& \\
\bottomrule
\end{tabular}
\end{table}

\begin{table}[H]
\centering

\begin{subtable}[t]{0.48\textwidth}
\vspace{0pt}
\centering
\caption*{$F$ derivatives}
\begin{tabular}{lccc}
\toprule
accuracy & 2 & 4 & 6 \\
\midrule
$c_1$ & $\mathbf{-0.29454}$ & $\mathbf{-0.294512}$ & $\mathbf{-0.294512}$ \\
$c_2$ & 0.918118 & 0.918023 & 0.918023 \\
$c_3$ & -1.65074 & -1.66183 & \\
$c_4$ & 11.3568 & 11.4265 & \\
\bottomrule
\end{tabular}
\end{subtable}
\hfill
\begin{subtable}[t]{0.48\textwidth}
\vspace{0pt}
\centering
\caption*{$c_1$ derivatives}
\begin{tabular}{lccc}
\toprule
accuracy & 2 & 4 & 6 \\
\midrule
$c_2$ & $\mathbf{0.918118}$ & $\mathbf{0.917928}$ & $\mathbf{0.917927}$ \\
$c_3$ & -1.6975 & -1.71308 & -8018.59 \\
$c_4$ & 11.3568 & 11.4614 & \\
\bottomrule
\end{tabular}
\end{subtable}

\end{table}

\begin{table}[H]
\centering

\begin{subtable}[t]{0.48\textwidth}
\vspace{0pt}
\centering
\caption*{$c_2$ derivatives}
\begin{tabular}{lccc}
\toprule
accuracy & 2 & 4 & 6 \\
\midrule
$c_3$ & $\mathbf{-1.60251}$ & $\mathbf{-1.60197}$ & $\mathbf{-1.60197}$ \\
$c_4$ & 11.1815 & 11.226 & 24998.9 \\
\bottomrule
\end{tabular}
\end{subtable}
\hfill
\begin{subtable}[t]{0.48\textwidth}
\vspace{0pt}
\centering
\caption*{$c_3$ derivatives}
\begin{tabular}{lccc}
\toprule
accuracy & 2 & 4 & 6 \\
\midrule
$c_4$ & $\mathbf{10.9004}$ & $\mathbf{10.8949}$ & $\mathbf{10.8949}$ \\
\bottomrule
\end{tabular}
\end{subtable}

\end{table}

\subsection{Momentum}

\begin{table}[H]
\centering
\caption*{Values of the momentum full counting statistics $F$ and charge cumulants $c_i$ as functions of $\lambda$.}
\begin{tabular}{c|ccccccc}
\toprule
$\lambda$
& $-0.03$ & $-0.02$ & $-0.01$ & $0$ & $0.01$ & $0.02$ & $0.03$ \\
\midrule
$F$
& -0.0107871
& -0.00727257
& -0.00367772
& 0
& 0.00376327
& 0.00761488
& 0.0115577 \\

$c_1$
& 0.347495
& 0.355406
& 0.363564
& $\mathbf{0.371979}$
& 0.380674
& 0.389648
& 0.398913 \\

$c_2$
& 0.779672
& 0.803881
& 0.829071
& $\mathbf{0.855291}$
& 0.88262
& 0.91109
& 0.94076 \\

$c_3$
& 2.37464
& 2.47034
& 2.57086
& $\mathbf{2.6765}$
& 2.78764
& 2.90455
& 3.02759 \\

$c_4$
& &
& &
$\mathbf{10.8331}$
& &
& \\
\bottomrule
\end{tabular}
\end{table}

\begin{table}[H]
\centering

\begin{subtable}[t]{0.48\textwidth}
\vspace{0pt}
\centering
\caption*{$F$ derivatives}
\begin{tabular}{lccc}
\toprule
accuracy & 2 & 4 & 6 \\
\midrule
$c_1$ & $\mathbf{0.372049}$ & $\mathbf{0.372004}$ & $\mathbf{0.372003}$ \\
$c_2$ & 0.85549 & 0.855396 & 0.855396 \\
$c_3$ & 2.73922 & 2.75298 & \\
$c_4$ & 11.2387 & 11.299 & \\
\bottomrule
\end{tabular}
\end{subtable}
\hfill
\begin{subtable}[t]{0.48\textwidth}
\vspace{0pt}
\centering
\caption*{$c_1$ derivatives}
\begin{tabular}{lccc}
\toprule
accuracy & 2 & 4 & 6 \\
\midrule
$c_2$ & $\mathbf{0.85549}$ & $\mathbf{0.855303}$ & $\mathbf{0.855302}$ \\
$c_3$ & 2.79841 & 2.81814 & 10128.9 \\
$c_4$ & 11.2387 & 11.3292 & \\
\bottomrule
\end{tabular}
\end{subtable}

\end{table}

\begin{table}[H]
\centering

\begin{subtable}[t]{0.48\textwidth}
\vspace{0pt}
\centering
\caption*{$c_2$ derivatives}
\begin{tabular}{lccc}
\toprule
accuracy & 2 & 4 & 6 \\
\midrule
$c_3$ & $\mathbf{2.67745}$ & $\mathbf{2.67653}$ & $\mathbf{2.67653}$ \\
$c_4$ & 11.0921 & 11.1319 & 23294.1 \\
\bottomrule
\end{tabular}
\end{subtable}
\hfill
\begin{subtable}[t]{0.48\textwidth}
\vspace{0pt}
\centering
\caption*{$c_3$ derivatives}
\begin{tabular}{lccc}
\toprule
accuracy & 2 & 4 & 6 \\
\midrule
$c_4$ & $\mathbf{10.8387}$ & $\mathbf{10.8332}$ & $\mathbf{10.8332}$ \\
\bottomrule
\end{tabular}
\end{subtable}

\end{table}

\end{document}